\documentclass{aa}  

\usepackage{graphicx}

\usepackage{txfonts}

\usepackage{natbib}
\bibpunct{(}{)}{;}{a}{}{,}

\usepackage[switch]{lineno}

\begin{document}
%\linenumbers

   \title{Accreted stars and stellar haloes of simulated galaxies in TNG50}

   \author{Bruno M. Celiz
          \inst{1,2,3}\fnmsep\thanks{\email{bruno.celiz@mi.unc.edu.ar}}
          \and
          Julio F. Navarro\inst{4}
          \and
          Mario G. Abadi\inst{2,3}
          }

   \institute{Facultad de Matemática, Astronomía, Física y Computación, UNC, Medina Allende s/n, X5000HUA, Córdoba, Argentina
        \and
            Instituto de Astronomía Teórica y Experimental, CONICET--UNC, Laprida 854, X5000BGR, Córdoba, Argentina
        \and
            Observatorio Astronómico de Córdoba, UNC, Laprida 854, X5000BGR, Córdoba, Argentina
        \and
            Department of Physics and Astronomy, University of Victoria, Victoria, BC, V8P 5C2, Canada}

   \date{Received XXX; accepted YYY}
 
    \abstract
    {We use the TNG50 cosmological hydrodynamic simulation to study the accreted stellar component and stellar haloes of isolated galaxies spanning a wide range of masses ($10^8<M_*/M_\odot<10^{11}$). We find that stars formed in the main progenitor (i.e., in-situ stars) typically dominate the inner regions as far as $\sim$10 half-light radii from the centre, implying that detecting uncontrovertible evidence for the presence of an accreted stellar halo requires probing the far outskirts of a galaxy. Stars from accreted, disrupted satellites (i.e., ex-situ stars) dominate beyond that radius (roughly $25\%$ of the virial radius, $r_{200}$), which we identify as the inner boundary of the outer stellar halo.  The fraction of accreted stars decreases monotonically with decreasing galaxy mass, $M_*$, from $\sim$20\% on average in $\sim$$2\times 10^{12}\, M_\odot$ haloes ($M_*\sim$$10^{11}\, M_\odot$) to $2$-$3\%$ in $\sim$$2\times 10^{10}\, M_\odot$ haloes ($M_*\sim$$10^{8}\, M_\odot$). The outer halo has a mass comparable to roughly $10\%$ of all accreted stars. Fewer than $\sim$30\% of stars in the outer halo are in-situ stars, many of which originate from star-forming satellites during the late stages of disruption, especially in low-mass systems. Accreted stars are systematically  more metal poor in less massive systems, which makes the outer haloes of dwarf galaxies a fertile hunting ground for extremely metal-poor stars. The density profile of accreted stars can be well approximated by a S\'ersic law, whose index, $n$ and effective radius, $R_{\rm eff}$, depend strongly on the total accreted mass. At given galaxy mass,  the more massive stellar haloes are systematically more concentrated (smaller $R_{\rm eff}$) and have steeper density profiles (larger $n$). The accreted component has generally larger $R_{\rm eff}$ than the main galaxy, although the two can have similar characteristic radii in the most massive systems with the largest accreted fractions. Our results provide a blueprint for interpreting observations of the outskirts of isolated galaxies in terms of their assembly histories. 
}

   \keywords{
   galaxies: dwarfs -- galaxies: formation -- galaxies: haloes -- galaxies: structure
    }

   \maketitle

\section{Introduction} \label{SecIntro}

In the standard $\Lambda$CDM cosmogony, galaxies grow hierarchically through the successive accretion of smaller systems \citep{WhiteRees1978,Frenk1988,Navarro1997}. Relics of the most massive accretion  events (mergers) are observed in the outskirts of massive galaxies, in the form of outer shells of stars \citep[see e.g.][]{MalinCarter81,MalinCarter83,Merritt2016,Monachesi2016, Rich2019,M-D2023}. The accretion of smaller galaxies typically lead to tidal disruption rather than  mergers, where the accreted stars form tidal tails and streams that, over time, can phase mix to form a featureless spheroidal component \citep{Johnston1996,Helmi1999,Helmi2003}. The combination of accretion and merger events leads to the formation of a faint stellar envelope known as the stellar halo \citep[e.g.][]{SearleZinn1978,BullockJohnston2005, Abadi2006}.

Significant progress has been made in recent decades identifying and characterizing stellar haloes and relics of individual accretion events, not only in Local Group galaxies such as the Milky Way (MW) and M31 \citep[see e.g.][]{Ibata2014,Helmi2018, Mateu2023}, but also in nearby MW-like galaxies \citep[see e.g.][]{Merritt2016,Monachesi2016, Rich2019,M-D2023}. Joint analysis of datasets such as those from the APOGEE \citep{Majewski2017} or GALAH \citep{GALAH2025} surveys, and \citet{Gaia2018} have revealed hundreds of stellar groups with coherent kinematics and chemical compositions, suggesting that they originate from  common progenitors \citep[see e.g.][]{FreemanBH2002,Belokurov2018, Koppelman2019}. 

Stellar haloes vary significantly from galaxy to galaxy. In the Milky-Way, the stellar halo is not particularly prominent and contributes only $\sim$1\% of its total luminosity \citep{Deason2019}, compared with M31, where the halo makes up $\sim$4\% of all stars \citep{McConnachie2009,Ibata2014}. M31's stellar halo is also more metal-rich, with a shallower and more extended mass density profile than the Milky Way's. These variations suggest that the accretion history of galaxies may differ significantly, even for  morphologically similar systems \citep[see e.g.][]{Amorisco2017,Harmsen2017,DsouzaBell2018}. 

Studies of MW-like galaxies outside the Local Group also reveal substantial galaxy-to-galaxy variations in stellar halo properties such as mass fraction, density profile, and age \citep{Merritt2016,Monachesi2016}, which are likely related to the stochastic nature of galaxy assembly histories \citep{BullockJohnston2005,DeLuciaHelmi2008,Cooper2010,Gomez2012,Proctor2024}.

Despite this stochasticity, some broad trends with galaxy stellar mass are expected. Because of the existence of a characteristic mass in the galaxy stellar mass ($M_*$) function (the 'knee') the total accreted mass, expressed as a fraction of $M_*$, should increase with $M_*$, especially for systems above the 'knee' \citep[see e.g.][]{Purcell2007}. It is less clear, however, what to expect in the dwarf galaxy regime. Do all dwarf galaxies have stellar haloes, or is there a characteristic galaxy mass below which all stars form in situ and accreted fractions become negligible? Well below the knee the scatter in the stellar mass-halo mass relation increases due to other baryonic processes and haloes may not experience mergers with other star-forming progenitors. On the other hand, questions also arise concerning the radial structure of stellar haloes: specifically, whether they are spatially more concentrated or extended than the main galaxy, and if a typical density profile adequately describes all stellar haloes or if they differ widely from system to system \citep{Abadi2006,Pill2014,Elias2018,Cooper2025}. How far from a galaxy should one probe to reach a region where the accreted component dominates over in-situ stars?

Another expected trend with galaxy mass concerns the average metallicity of the halo. Because of the mass-metallicity relation, accreted stellar haloes are expected to be in general more metal-poor than their host galaxies \citep[see e.g.][]{Lee2006, Kirby2013}. This is certainly true in most well-studied systems, such as the Milky Way, M31, and other MW-like galaxies \citep[see e.g.][]{Tissera2012, Harmsen2017, DsouzaBell2018}, but it is unclear how these trends extrapolate to the regime of dwarf galaxies, which tend to be metal-poor themselves. How metal-poor are haloes expected to be in that case? Similar questions may be posed regarding the relative age of stars in the main galaxy and its halo. Although it is reasonable to expect that haloes will be older than the galaxy host, it is unclear what to expect when comparing the halo of a dwarf galaxy with that of a MW-like galaxy, or even a more massive system. Which halo is expected to be the oldest?

In this work, we have addressed some of these questions using the cosmological hydrodynamical simulation TNG50 \citep[][]{Nelson2019TNG50, Pillepich2019}. Previous work has already used Illustris and IllustrisTNG to study accreted mass and stellar haloes but they have so far focused on massive galaxies \citep[see e.g.][]{Pill2014, Pillepich2018,R-G2016,Elias2018}. These simulations have shown that: (i) the spherically averaged stellar density profile at large galactocentric distances is well approximated by a power-law, $\rho_{*}(r) \propto r^{\alpha}$ with $-7 \lesssim \alpha \lesssim -2$; (ii) steeper halo profiles are associated with galaxies that have undergone early major mergers; and (iii) the stellar halo mass fraction correlates with assembly time and galaxy morphology, such that galaxies with relatively more massive stellar haloes assemble later and are less rotationally supported.

While these results are broadly consistent\footnote{see, however, \citet{Merritt2020} for a different view.} with observed properties of stellar haloes of massive galaxies \citep[see e.g.][]{Elias2018}, the extension to less massive systems, as discussed above is yet to be fully understood \citep{Carlin2019, Deason2022, Fielder2024, Tau2024, Cooper2025,Gonzalez-Jara2025}, and is one of the main goals of our study. This is important, especially in anticipation of upcoming data from, e.g., the LSST \citep{LSST2019}, Euclid \citep{Euclid2024} and Roman \citep{WFIRST2019} surveys.

This paper is organized as follows. In Section \ref{SecMethods} we describe the simulation used, the stellar halo and accreted mass definition, and an illustrative example of a dwarf galaxy. In Section \ref{SecResults} we extend the analysis to the resulting galaxy sample, comparing their accreted and stellar halo component and different masses at redshift $z=0$. We then characterize the spatial distribution of accreted stars study their metallicities. Finally, in Section \ref{SecConc} we summarize our results and present our conclusions.

\section{Numerical Methods} \label{SecMethods}

\subsection{The TNG50 simulation} \label{SecSims}

Our analysis uses The Next Generation Illustris simulations \citep[IllustrisTNG\footnote{\url{https://www.tng-project.org/}},][]{Miranacci2018, Naiman2018, Pillepich2018modelTNG,Springel2018,Nelson2019}. This suite of magneto-hydrodynamic cosmological simulations evolve cosmologically representative volumes in a $\Lambda$CDM Universe with parameters consistent with \citet[][$h = 0.6774;~\Omega_m = 0.3089;~\sigma_8 = 0.8159 $]{Planck2016}. 
The initial conditions were set at redshift $z=127$ and were obtained
with the code \texttt{N-GENIC} \citep{Springel2005}. The simulations are evolved with the moving mesh code \texttt{AREPO} \citep{Springel2010, Pakmor2016} forward in time until $z=0$. Detailed properties of dark matter and baryonic particles were systematically recorded in 100 snapshots, spaced spaced by time intervals of $\sim$0.15 Gyr.

In particular, we focused on the TNG50-1 run of the suite \citep[][TNG50 hereafter]{Nelson2019TNG50, Pillepich2019}. This simulation is the highest-resolution run of a $51.7^3~\mathrm{Mpc^3}$ periodic box and includes $2160^3$ particles of dark matter, each with a mass $m_{\mathrm{DM}} = 4.5 \times 10^5~M_{\mathrm{\odot}}$, alongside an equivalent number of gas cells, targeting a baryonic mass of $m_{\mathrm{baryon}} = 8.5 \times 10^4~M_{\mathrm{\odot}}$. The Plummer-equivalent gravitational softening length for dark matter and stars was set at $\epsilon_{\mathrm{DM,*}} = 0.29$ kpc, and for adaptive gas cells a minimum value of $\epsilon_{\mathrm{gas}} = 0.07$ kpc (both at redshift $z=0$).

The TNG50 simulation incorporates a baryonic treatment based on the earlier Illustris project \citep[][]{Vogelsberger2013,Vogelsberger2014,Weinberger2017,Pillepich2018modelTNG}. In TNG50, gas is allowed to cool down to a temperature of $T = 10^4$ K, with cooling and heating rates calculated based on local density, redshift, and metallicity. To prevent artificial fragmentation, gas exceeding a density of $n = 0.13~\mathrm{cm^{-3}}$ was modelled using a two-phase equation of state \citep{SpringelHernquist2003}. Star formation occurs in gas cells above this density threshold, and stellar
particles are born assuming a Chabrier initial mass function \citep{Chabrier2003}, inheriting the cell’s mass, momentum, and
metallicity. Their subsequent stellar evolution and feedback implemented following the prescriptions described in \citet{Pillepich2018modelTNG}, where different stellar populations evolve and return mass and metals to the interstellar medium.

In this simulation, stars enrich their surroundings with elements through stellar winds and supernovae. This happens by injecting metal-rich gas into nearby cells, effectively mixing newly ejected metals from a star particle into the surrounding gas, proportionally reducing the particle mass while its metallicity remains fixed, ensuring that a star particle is never destroyed. In addition, TNG includes gas radiative processes, black hole formation and evolution, and black hole feedback in two modes: a thermal mode (activated when the black hole is in a high accretion rate state) and a kinetic mode (that occurs when it is in a low accretion rate state), as described in \citet{Weinberger2017}.

The available group and galaxy catalogues were generated using the friends-of-friends \citep[{\texttt FoF},][]{Davies1985} and \texttt{SUBFIND} \citep{Springel2001, Dolag2009} algorithms. We used the \texttt{SubLink} \citep{R-G2015} merger trees to track the temporal evolution of galaxies.

\subsection{Simulated galaxy sample}
\label{SecSimGx}

Our analysis concentrates solely on isolated galaxies, defined as the central galaxies of their respective FoF groups. We include systems with  virial\footnote{Throughout this paper, virial quantities are defined at the radius enclosing 200 times the critical density for closure.} mass $\log(M_{200}/M_{\mathrm{\odot}}) > 10.3$, spanning a wide range of galaxy luminosities, from dwarf galaxies to  massive systems. This selection implies that our simulated galaxy sample is essentially complete for $\log(M_{*}/M_{\mathrm{\odot}}) > 8.0$ (see Section \ref{sec:AccMassFrac}). 

To exclude ongoing merger events and systems clearly out of equilibrium, we filtered out galaxies with a satellite whose stellar mass exceeds $10$\% that of its host. We also used the merger trees to exclude backsplash systems (i.e., those that were previously associated with a larger group or cluster at some point in their history). For each galaxy, we considered all dark matter particles, stellar particles, and gas cells gravitationally bound within its virial radius $r_{200}$, excluding satellites (defined as stars bound to another \texttt{SUBFIND} subhalo). All distances and sizes shown throughout this work are in physical distance units.

\subsection{In-situ vs accreted (ex-situ) stars} \label{SecInExSitu}

\begin{figure}
    \centering
    \includegraphics[width=\columnwidth]{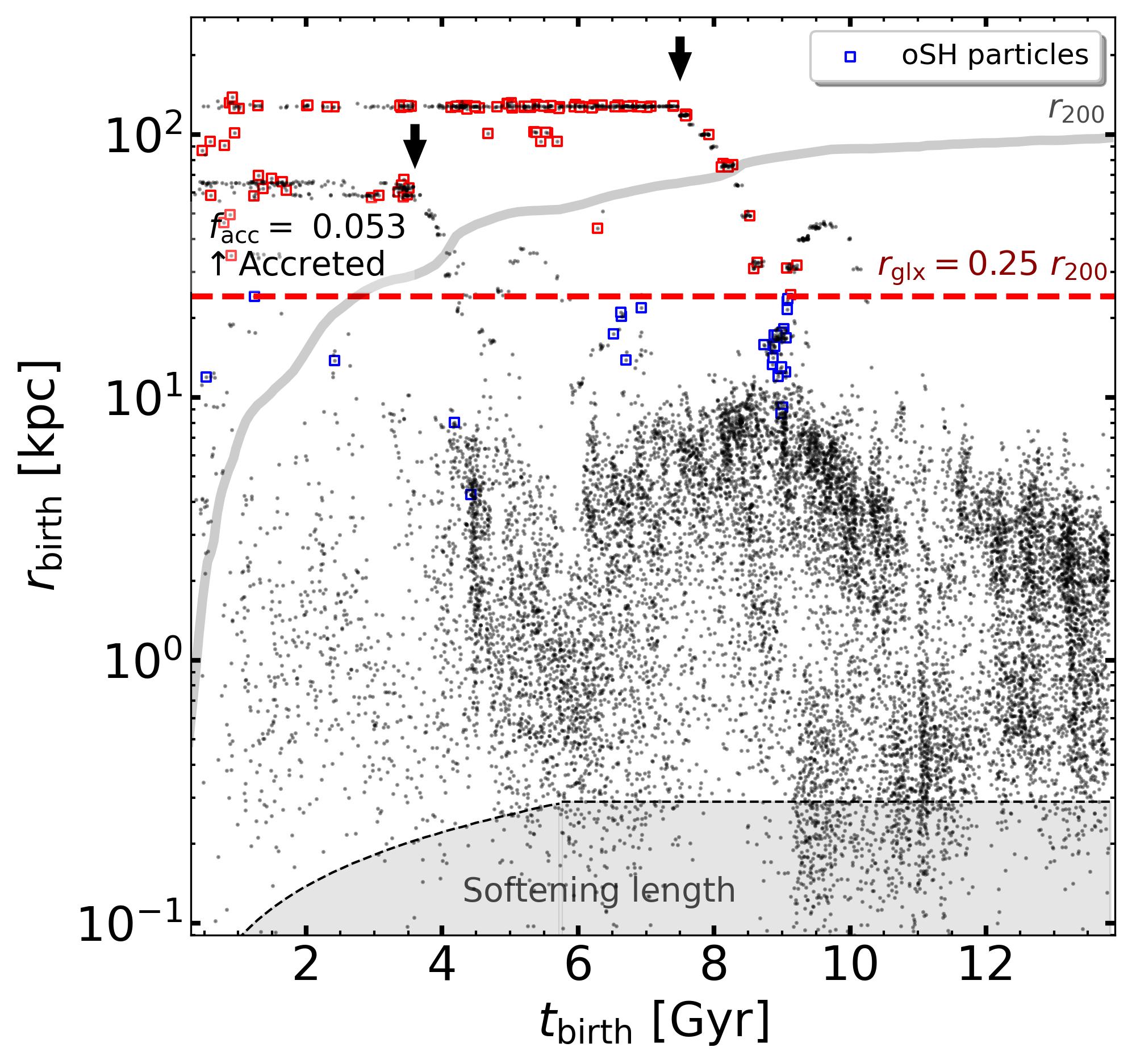}
    
    \caption{Distance to the main progenitor at formation time, $r_{\rm birth}$, as function of cosmic formation time, $t_{\rm birth}$, for all star particles (black dots) bound to the example galaxy TNG50-714463 (see Fig. \ref{fig:illustrative_example_insitu_exsitu}). Star particles born within the main progenitor (i.e., $r_{\rm birth} < r_{\rm glx}$) are classified as in-situ; those with $r_{\rm birth} > r_{\rm glx}$ (above the threshold radius shown as a red horizontal dashed line) are defined as ex-situ, or 'accreted'. The example galaxy accreted two satellites at $t \approx 3.7$ Gyr and $t \approx 7.5$ Gyr (indicated with black arrows). Stars born in these satellites contribute $\log(M_{\mathrm{acc}}/M_{\mathrm{\odot}}) = 7.68$ to the host galaxy, for a total accreted mass fraction $f_{\rm acc}=0.053$. Stars in the outer stellar halo at $z=0$ ($r > r_{\rm glx}$, labelled 'oSH') are highlighted with empty  squares, blue if they are in-situ, and red if they are accreted (see Fig.~\ref{fig:illustrative_example_insitu_exsitu}). The gray shaded region denotes distances smaller than time-evolving force-softening length for stars and dark matter, $\epsilon_{\mathrm{DM,*}}$.}
    \label{fig:illustrative_example_tbirth_rbirth}
\end{figure}

\begin{figure*}
    \centering
    \includegraphics[width=\textwidth]{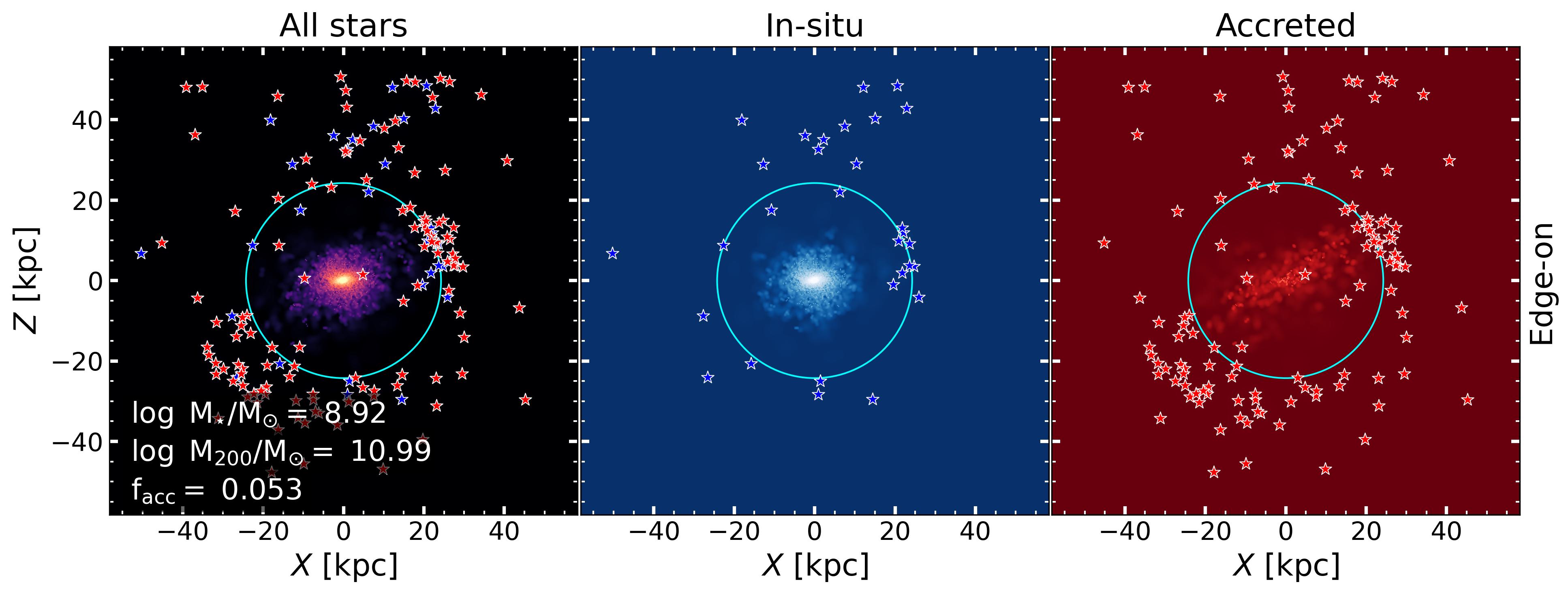}
    \caption{Edge-on projection of a simulated dwarf galaxy (TNG50-714463, stellar and virial masses are shown in the legend, left panel), split into its in-situ (centre panel) and accreted (right panel) stellar components. Our definition of galaxy radius, $r_{\rm glx} = 0.25 ~ r_{200}= 24.25$ kpc, is indicated by the cyan circles, separating the main galaxy from its outer stellar halo. This example has an accreted mass comparable to the median value for  galaxies  of comparable mass ($f_{\rm acc} = 0.053$). Brighter colours indicate higher number density of star particles, while  starred symbols are outer halo stars coloured according to their origin (blue for in-situ; red for accreted). Images generated with PySPHViewer \citep{BenitezLlambay2017}.}
    \label{fig:illustrative_example_insitu_exsitu}
\end{figure*}

To characterize the accreted stellar component of a galaxy we first distinguish between stars formed in the main progenitor (i.e., in-situ) and accreted star particles contributed by disrupted satellites (i.e., ex-situ). Our definition is based on the at-birth distance of a star to the main progenitor of the galaxy ($r_{\mathrm{birth}}$). Since the simulation outputs are discretized into $100$ snapshots, birth distances are measured using the first snapshot after the particle's birth time, $t_{\rm birth}$, i.e. $r_{\mathrm{birth}} = |\Vec{x}_{\mathrm{*}}(t_{\rm birth}) - \Vec{x}_{\mathrm{glx}}(t_{\rm birth})| $, where $\Vec{x}_{\mathrm{glx}}$ is the position of the centre of the main progenitor (defined by the most bound particle of the subhalo) and $\Vec{x}_{*}$, the position of a newly born star particle.

As an example, Fig.~\ref{fig:illustrative_example_tbirth_rbirth}  shows, as a function of birth time, $r_{\rm birth}$ for every star particle bound at $z=0$ to TNG50-714463, a dwarf galaxy with $\log(M_{*}/M_{\mathrm{\odot}}) = 8.92$ and virial mass $\log(M_{200}/M_{\mathrm{\odot}}) = 10.99$. 

Accreted stars are easily identified in this panel; the majority are contributed by two star-forming satellites, one which first crosses the virial radius of the main progenitor (shown by the grey solid curve) at $t \sim 4$ Gyr, and another at $t \sim 8$ Gyr. The horizontal portion of the tracks are an artifact due to the fact that we can only compute $r_{\rm birth}$ once a star is in the same FoF group as the main progenitor. For ex-situ stars formed before a satellite joins the main progenitor's FoF, we define $r_{\rm birth}$ as the distance at that particular time, indicated with downward arrows in the figure.

Based on this figure, we identify as 'ex-situ' those stars that formed outside a fixed fiducial 'galaxy radius', which we define as 25\% of the virial radius at $z=0$ ($r_{\rm glx}=0.25\, r_{200}$). This choice is, of course, arbitrary, but we shall see below that it roughly delineates the boundary of the regions where in-situ or ex-situ stars dominate at present.

In this example, the large majority of stars are formed in-situ, with ex-situ stars contributing only $5.3\%$ of $M_*$. Note also that the distinction between in-situ and ex-situ stars is inevitably fuzzy, as satellites may form stars inside and outside $r_{\rm glx}$ as they orbit the main progenitor. In this case, a fraction of stars born in satellites would be classified as in-situ \citep[other authors classify these particles as 'endo-debris', see e.g.][]{Tissera2013,Tissera2014,Gonzalez-Jara2025}. In general, these contribute only a small fraction of the total \citep[see; e.g.][for further discussion]{Cooper2015,R-G2015}.

Our choice of a fixed physical radius to define the in-situ/ex-situ components differs from other choices in the literature, which are typically based on merger trees, where in-situ stars are identified with those bound at birth to the main progenitor. This choice can be problematic, as it may designate as 'accreted' many  stars that form in tightly-bound high-redshift clumps which precede the emergence of the true main progenitor of the system. At those times, it is difficult to ascertain which of those early clumps is the true main progenitor, and merger trees often choose a main branch that does not coincide with the progenitor where most stars form at the time. Without some kind of correction, merger tree-based methods might therefore spuriously designate many of those early-forming stars as 'accreted', although they form in close physical proximity to in-situ ones and largely remain today confined to within the inner regions of the galaxy. Defining accretion by using $r_{\rm glx}$, as we have chosen to do, tries to circumvent such problem by  identifying as 'accreted' mainly stars formed physically far from the main progenitor. The choice of a fixed $r_{\rm glx}$ is, although imperfect, simple and effective in this regard.

\subsection{Dwarf galaxy stellar halo: an example} \label{SubsecExample}

As an illustrative example, we show in Fig. \ref{fig:illustrative_example_insitu_exsitu} the edge-on\footnote{We use the angular momentum of the youngest stars (with ages less than 1 Gyr) to define the $z$-axis of the projection.} projection of the  dwarf galaxy TNG50-714463 shown in Fig.~\ref{fig:illustrative_example_tbirth_rbirth}.  We deliberately choose a low-mass galaxy so that we can highlight individual stars in the stellar halo without overlap/confusion. 

We decompose the total stellar component (left panel) by origin: stars born in the main progenitor (in-situ, blue, central panel) or accreted (ex-situ, red, right panel).  Cyan circles indicate $r_{\mathrm{glx}} = 0.25 ~ r_{200}$, used as the distance threshold at birth for labelling particles as in-situ or ex-situ. Star particles found at $z=0$ outside $r_{\mathrm{glx}}$ are shown with starred symbols, coloured by their origin. The total accreted stellar mass\footnote{Measured as the sum of the mass of all star particles classified as accreted, within the virial radius.} of this galaxy is $\log(M_{\mathrm{*,acc}}/M_{\mathrm{\odot}}) = 7.68$, which implies an accreted mass fraction of $ f_{\mathrm{acc}} =M_{\mathrm{*,acc}}/M_{*} \approx 0.053$, similar to the median value for galaxies of similar mass (see Section \ref{sec:AccMassFrac}). Interestingly, the spatial distribution of the accreted component is misaligned with the disc of the main galaxy, and reflects the orbital plane of the accretion event that contributed the majority of accreted stars.

\section{Results} \label{SecResults}

\subsection{The outer stellar halo} \label{SecInSituRadius}

\begin{figure}
    \centering
    \includegraphics[width=0.8\columnwidth]{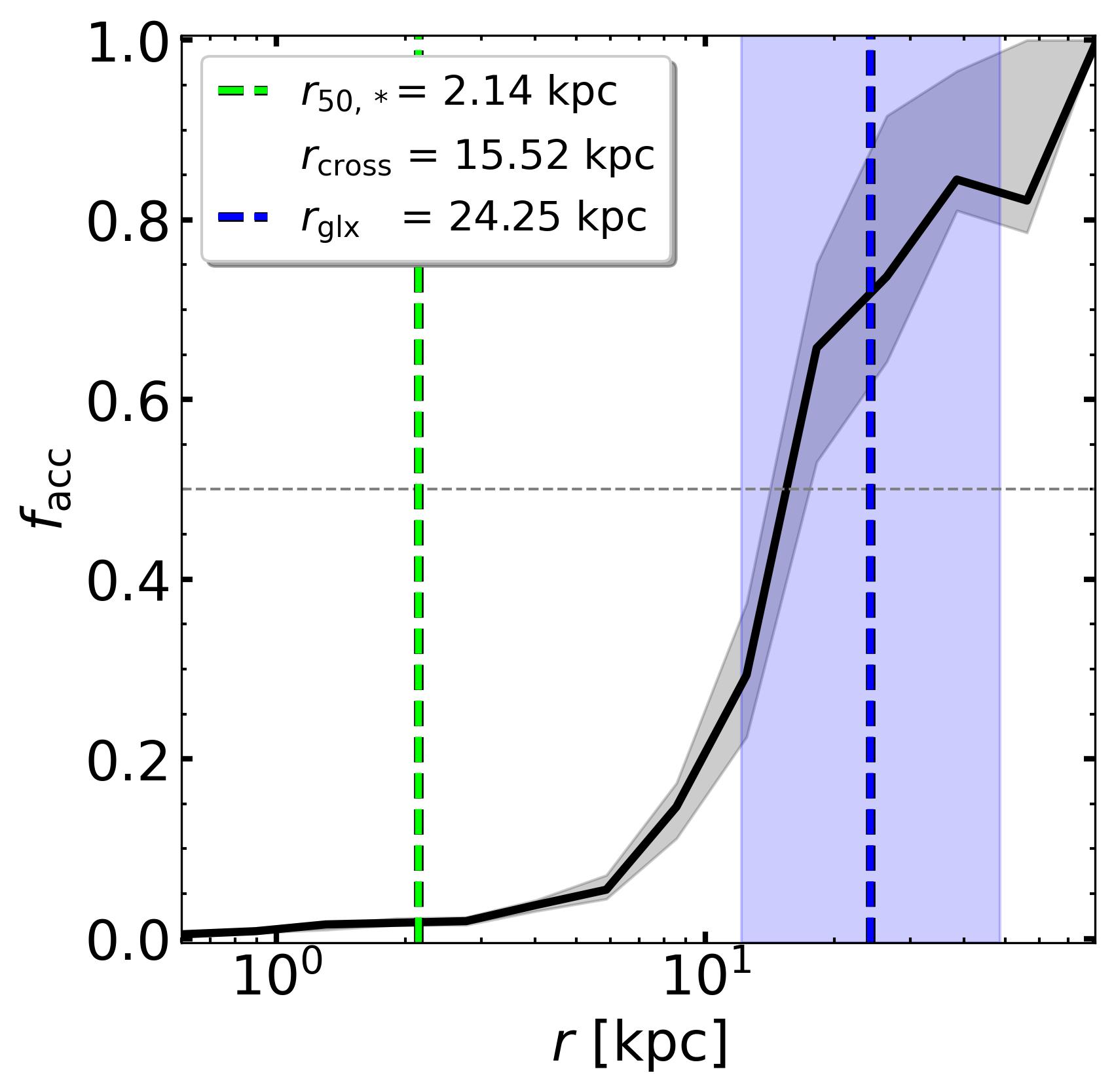}
    \caption{Accreted stellar mass fraction in 3D radial bins, $f_{\rm acc}(r)=M_{\rm acc}(r)/M_{*}(r)$, for the illustrative example galaxy TNG50-714463. The vertical green dashed line indicates the stellar half-mass radius of the galaxy. Accreted stars begin to dominate the stellar mass ($f_{\rm acc} > 0.5$) at $r \gtrsim 15.5$ kpc. This defines the 'crossing radius', $r_{\rm cross}$, of the galaxy. This radius is more than 7 times larger than the stellar half-mass radius of this galaxy, and is smaller than our adopted definition of 'galaxy radius' $r_{\rm glx}$, shown with a blue dashed line. Changes of $r_{\rm glx}$ by a factor of 2 (blue shaded regions) produce only small variations in the accreted fraction profile (grey shaded regions), with $r_{\rm cross} = 15 \pm 1$ kpc.}
    \label{fig:illustrative_example_profiles}
\end{figure}

\begin{figure*}
    \centering
    \includegraphics[width=\textwidth]{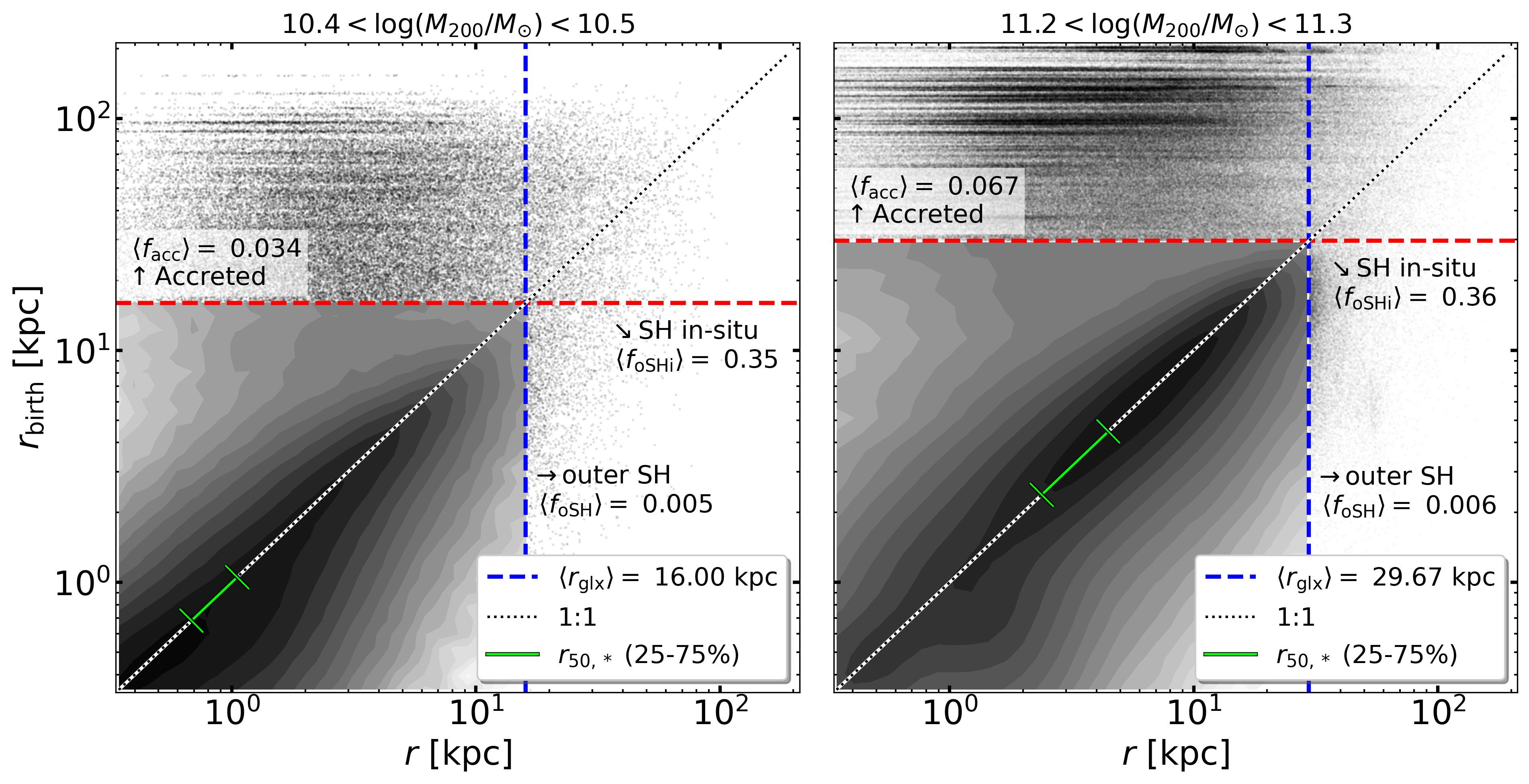}
    
    \caption{Distance to the main progenitor at birth, $r_{\mathrm{birth}}$, vs. galactocentric distance at present time, $r$, of all star particles (excluding satellites) for 773 galaxies with virial mass $10.4 < \log(M_{200}/M_{\odot}) < 10.5$ (left) and 165 galaxies with virial mass $11.2 < \log(M_{200}/M_{\odot}) < 11.3$ (right), stacked. The  dotted white diagonal line shows the 1:1 relation, i.e., stars at the same galactocentric distance as when they were born. Blue dashed vertical line indicates the median $r_{\mathrm{glx}}$, which separates the main galaxy from the outskirts (the outer stellar halo). Stars born at  $r_{\rm birth} > r_{\mathrm{glx}}$ (above the red dashed horizontal line) are classified as accreted. Most in-situ stars lie close to the 1:1 line, and most accreted stars end up in the inner regions of the galaxies. Consequently, the mass of stars in the stellar halo is one order of magnitude smaller than the accreted mass fraction. The 25-75th percentile of the half-mass stellar radius of the stacks of galaxies are shown with a green interval over the 1:1 line.}
    \label{fig:r_birth_rz0_2bins}
\end{figure*}

\begin{figure}
    \centering
    \includegraphics[width=\columnwidth]{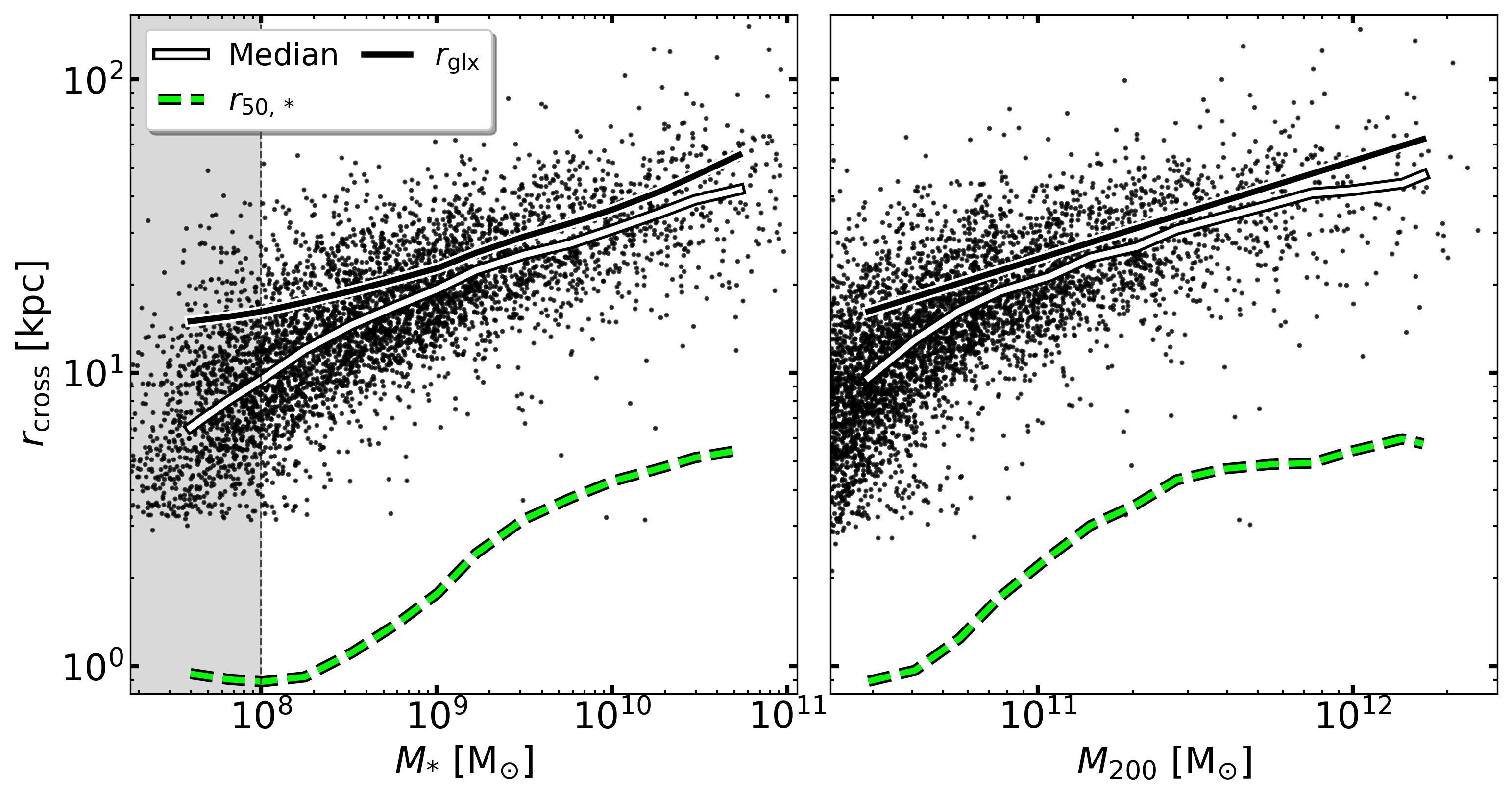}
    \caption{Distance at which  accreted stars begin to dominate, $r_{\rm cross}$, for all galaxies  in our sample (black dots) as a function of stellar mass (left panel) and virial mass (right panel). For comparison, we show its median (white solid line), the median stellar half-mass radius $r_{50,*}$ (green dashed line), and that of the adopted galaxy radius $r_{\rm glx}$ (black solid line). The typical galactocentric distance beyond which most stars are accreted is roughly $\sim$8 times larger than the stellar half-mass radius, and close to the galaxy radius definition, $ r_{\rm glx}=0.25\, r_{200}$.}
    \label{fig:r_glx_sample}
\end{figure}

Simulations have revealed that stellar haloes can contain a significant fraction of in-situ stars \citep[see e.g.][]{Tissera2012,Tissera2013,Cooper2015,Monachesi2019,Tau2024,Gonzalez-Jara2025}. Some of these stars may have been %pushed out
pulled into the halo by accretion events or mergers, while others may have been born directly in the halo out of gas instabilities in the circumgalactic medium \citep[see e.g.,][]{Ahvazi2024}.

Regardless of their origin, this in-situ 'contamination' complicates the interpretation of outer stellar envelopes as repositories of mainly accreted mass.  Consequently, various authors have explored different choices for the radius beyond which accreted stars dominate, such as multiples of the stellar half-mass radius, or a surface brightness density limit, or a fixed physical distance at $z=0$ \citep[see e.g.][]{DSouza2014, Pill2014, Pillepich2018, Monachesi2016, Monachesi2019, Merritt2016, Merritt2020, Harmsen2017}. 

We examine this issue in Fig.~\ref{fig:illustrative_example_profiles} for the  example galaxy TNG50-714463. This figure shows, as a function of 3D radius, the accreted mass fraction measured in $20$ equally-spaced logarithmic radial shells. The accreted mass fraction increases with galactocentric distance, from $f_{\rm acc} < 0.05$ within the 3D stellar half-mass radius ($r_{50,*}$, green dashed line) to $f_{\rm acc} = 0.5$ at $r \approx 15 \, {\rm kpc}$. Outside this radius, often referred to as 'crossing radius' \citep[$r_{\rm cross}$, see][]{R-G2016}, the accreted component dominates. The crossing radius definition is quite robust to changes in $r_{\rm glx}$: the grey-shaded region in Fig.~\ref{fig:illustrative_example_profiles} shows how the radial profile of the accreted fraction changes when varying $r_{\rm glx}$ by a factor of 2. Note that in this case the accreted component only dominates outside $r_{\rm cross} \sim 7 ~ r_{50,*}$, a distance greater than typically assumed to define the stellar halo of a galaxy \citep[see e.g.][]{Elias2018,Monachesi2019,Tau2024}, and closer to the here adopted 'galaxy radius' definition ($r_{\rm glx}$, blue dashed line). 

Fig. \ref{fig:r_birth_rz0_2bins} indicates that this is generally true for most galaxies in our TNG50 sample. Here we show 
the birth radius ($r_{\mathrm{birth}}$) versus the present-day galactocentric distance of all star particles in systems within two narrow bins of virial mass: $773$ galaxies with  $10.4 < \log(M_{200}/M_{\odot}) < 10.5$ ($M_*\sim 10^8\, M_\odot$, left panel) and $165$ galaxies with $11.2 < \log(M_{200}/M_{\odot}) < 11.3$ ($M_*\sim 10^{10}\, M_\odot$, right panel). Regions with large numbers of points, close to the innermost region of galaxies, are shown as filled contours.

The high density of stellar particles around the 1:1 line represents the bulk of the in-situ component, and indicate that most in-situ stars stay, on average, at distances similar to where they formed. This is not unexpected in isolated galaxies, where stars experience little orbital heating or secular evolution unless perturbed by tidal forces from minor or major mergers \citep{B-L2016,Navarro2018, Graus2019, Mercado2021}. 

The size of the in-situ-dominated region increases with mass, from $\sim$10 kpc for the low-mass systems in the left panel of Fig.~\ref{fig:r_birth_rz0_2bins} to $\sim 30$ kpc for the more massive systems in the right-hand panel. Further scrutiny shows that the condition $r<r_{\mathrm{glx}}$ describes this region well for galaxies of all masses. This boundary is shown by the vertical blue dashed lines in the same figure, and is typically of order $7$-$10\times$ the half-mass radius\footnote{We also note that the stellar half-mass radius of dwarf galaxies in TNG50 is roughly constant at $\sim 1$ kpc \citep[see e.g.][]{Almeida2024, Martin2025}, likely an artifact of limited numerical resolution, as discussed in detail by \citet{Celiz2025}.} of the stars (shown by the green segments on the 1:1 line).

Fig.~\ref{fig:r_birth_rz0_2bins} also shows that accreted stars (i.e., those above the red horizontal dashed line) span a wide range of radii at $z=0$, with the majority of them well inside $r_{\rm glx}$. Some in-situ stars do spill outside this radius, but they contribute less than 40\% of the total. The fraction of stars in each region of this panel is listed in the figure legend.

Finally, we compare the size of the in-situ-dominated region (as measured by $r_{\rm cross}$) with the stellar half-mass radius, $r_{50,*}$, and the galaxy radius, $r_{\rm glx}$ as a function of $M_*$, as shown in Fig.~\ref{fig:r_glx_sample}. The white curves trace the median $r_{\rm cross}$, which compares well with the average $r_{\rm glx}$, shown by the black curves. The stellar half-mass radius of the galaxy is roughly $7$-$10\times$ smaller than $r_{\rm cross}$, implying that only studies of the far outskirts of a galaxy are able to probe primarily accreted stars, with little contamination from in-situ stars.

We note that the $r_{\rm cross}$ values obtained are robust to the definition of galaxy size ($r_{\rm glx}$), which we use to define the outer stellar halo and classify accreted stars. For instance, changing the threshold distance for ex-situ classification, $r_{\rm glx}$, by a factor of 2 only slightly modifies the radius where the accreted component begins to dominate the stellar mass, as the example in Fig. \ref{fig:illustrative_example_profiles} illustrates.

\subsection{Accreted mass fractions}\label{sec:AccMassFrac}

\begin{figure}
    \centering
    \includegraphics[width=\columnwidth]{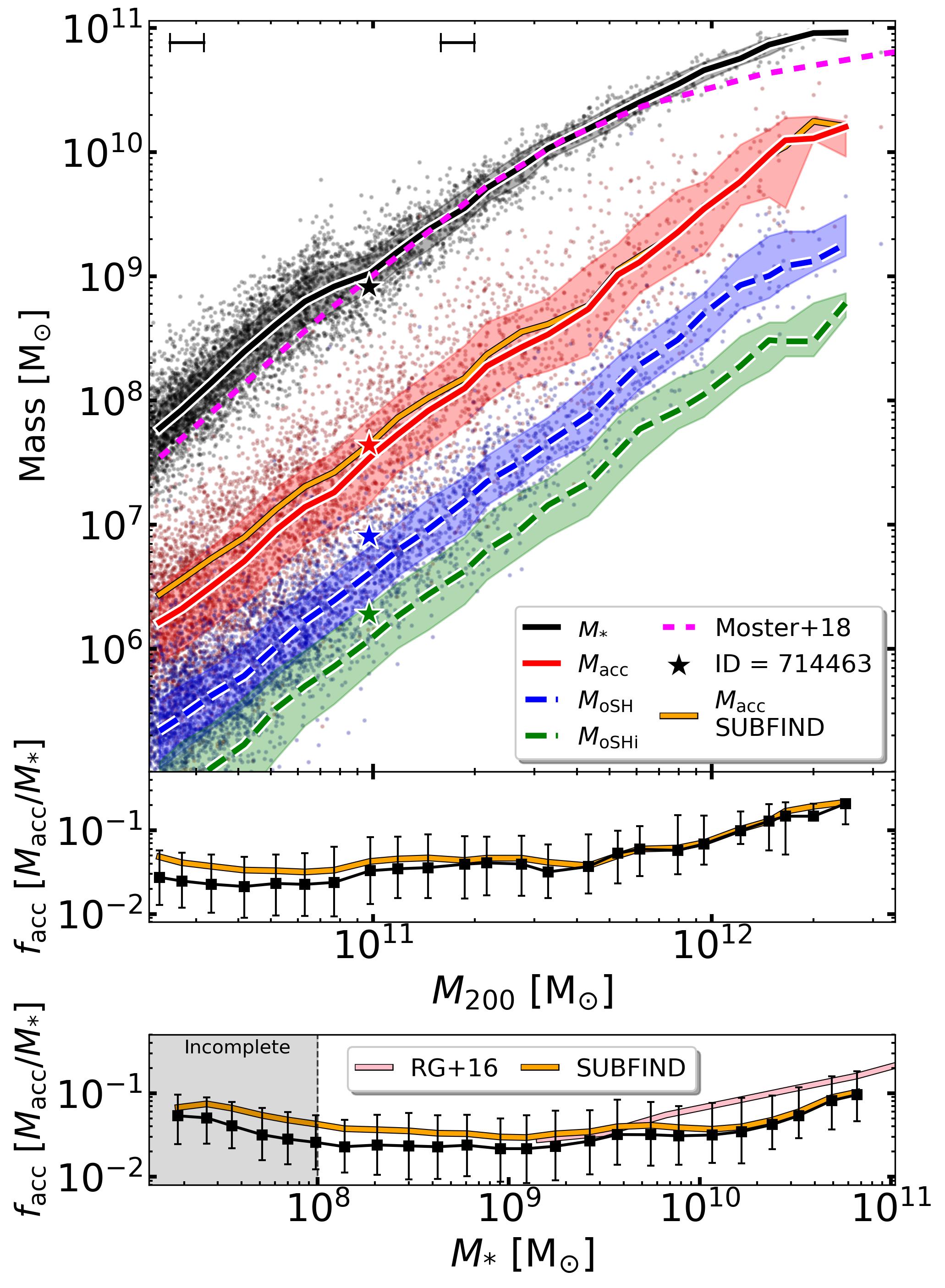}
    \caption{Main panel: Stellar mass (black), accreted mass (red), outer stellar halo mass (blue) and in-situ mass in the stellar halo (green) as a function of virial mass for the 5131 isolated galaxies (dots) in our sample. Solid lines and shaded regions show the median trend and the 25-75th percentile regions. The dashed magenta line shows the \citet{Moster2018} abundance-matching relation at $z=0$, whereas the solid orange line shows the accreted mass obtained when using publicly available values from a subhalo membership method, based on \texttt{SUBFIND}. The two black segments at the top indicate the range of mass of the stacks shown in Fig. \ref{fig:r_birth_rz0_2bins}. The middle panel shows the accreted-to-total stellar mass fraction, $f_{\rm acc} = M_{\rm acc}/M_{*}$, as a function of virial mass, and the bottom panel $f_{\rm acc}$ as function of galaxy stellar mass. The incompleteness of the sample at stellar masses $M_* < 10^{8} ~ M_{\odot}$, generates a spurious upturn in the trend. With a pink line we show the results reported by \citet{R-G2016} for the Illustris simulation.}
    \label{fig:Mvir_Mstar_Macc_MSH}
\end{figure}

We use the definitions presented in the preceding sections to compute, at $z=0$, for each galaxy in our sample, the total stellar mass of the galaxy, $M_{*}$, the accreted stellar mass, $M_{\mathrm{acc}}$, the stellar mass in the outer halo, $M_{\mathrm{oSH}}$, and the in-situ stellar mass in the outer stellar halo $M_{\rm oSHi}$, defined as follows:
\begin{equation}
\begin{split}
    M_{*} &= M_{*}(r < r_{200}) \\
    M_{\rm acc} &= M_{*}(r_{\rm birth} > r_{\mathrm{glx}}) \\
    M_{\rm oSH} &= M_{*}(r > r_{\mathrm{glx}}) \\
    M_{\rm oSHi} &= M_{*}(r > r_{\mathrm{glx}} \ \& \ r_{\rm birth}<r_{\rm glx}) \\
\end{split}
\label{eq:Mass_defs}
\end{equation}
where $r_{\mathrm{glx}} = 0.25 ~ r_{200}$ at $z=0$, and the definitions exclude %all satellites. 
stars bound to other subhaloes of the main halo. Fig.~\ref{fig:Mvir_Mstar_Macc_MSH}  shows these masses for all galaxies as function of virial mass. The $M_*$-$M_{200}$ relation (black solid line) agrees well with abundance matching model predictions \citep[see e.g.][shown with a dashed magenta line]{Moster2018}. However, TNG50 galaxies with $M_{200} < 10^{11} ~ M_{\odot}$ exhibit slightly higher total stellar masses, caused by a numerical artifact at the innermost region of galaxies that increase star formation in some dwarf galaxies, as discussed in \citet{Celiz2025}. Shaded regions highlight the 25-75th percentile range of each relation. The starred symbols indicate the TNG50-714463 galaxy used as an illustrative example in Section \ref{SubsecExample}.

Our sample, by definition, contains all central galaxies with $10.3 < \log(M_{200}/M_{\odot}) < 12.3$ and is complete for stellar masses $\log(M_{*}/M_{\odot}) > 8.0$. The scatter in accreted masses is fairly large, with a 25-75th percentile range that spans a factor of $\sim$5 on average. Most of the accreted mass resides inside $r_{\rm glx}$, as may be seen from the vertical distance between the blue and red curves; typically the mass in the outer stellar halo makes up only $10\%$ of the total accreted mass. The outer halo also contains some in-situ stars, but they make up typically fewer than $30\%$ of all outer halo stars.

The bottom panels of  Fig.~\ref{fig:Mvir_Mstar_Macc_MSH} show, as a function of either $M_{200}$ or $M_*$, the accreted stellar mass fraction, $f_{\rm acc}=M_{\rm acc}/M_*$. At large masses, $f_{\rm acc}$ correlates well with halo or galaxy mass, but it seems to converge to a low but constant value of roughly $2\%$ at low masses, with substantial scatter. Note the slight upturn in $f_{\rm acc}$ at low $M_*$, which is caused mainly by the fact that our sample is selected with a sharp lower bound in $M_{200}$, and so it is incomplete and biased for $M_*<10^8\, M_\odot$. 

As anticipated when discussing our choice of $r_{\rm glx}$ (Sec.~\ref{SecInExSitu}), our accreted mass fractions are lower than those obtained using a  subhalo membership criterion\footnote{Available as the \texttt{AccretedStellarMass\_SinceRedshift5} values found in the \texttt{Merger History} public catalogue from the IllustrisTNG webpage: \url{https://www.tng-project.org/data/docs/specifications/#sec5y}.} to classify in-situ/ex-situ stars. This is shown by the orange solid lines in the bottom panels of Fig.~\ref{fig:Mvir_Mstar_Macc_MSH}. The same panels show as well the results reported by \citet[]['RG+16']{R-G2016} for the TNG100 simulation with a pink solid line, both broadly in agreement with our values reported here over the relevant range of masses.

It isn't immediately clear why $f_{\rm acc}$ converges to $\sim 2\%$ for $M_* < 10^9 \, M_\odot$. In particular, it would be interesting to see whether fainter dwarfs actually lack an accreted component altogether. Our analysis cannot rule this out, but suggests that 'halo-free' dwarfs, if present, have stellar masses below $10^8\, M_\odot$ \citep[see Section 4 of][for further discussion]{Cooper2025}. Indeed, of the 5131 isolated centrals in our sample, only 6 galaxies lack accreted stars ($M_{\rm acc}=0$), and only $84$ do not exhibit a stellar halo ($M_{\mathrm{SH}} = 0$), all with stellar masses $\log(M_{*}/M_{\odot}) < 8.5$.
% B: Acá podría agregar algo de resolución? Tipo, "We note, however, that the progenitors of such galaxies are poorly resolved in the simulation, showing typically larger sizes and having a poorly constrained M*-M200 relation."

\subsection{Projected density profile of accreted stars} \label{SecHaloDensProf}

\begin{figure}
    \centering
    \includegraphics[width=\columnwidth]{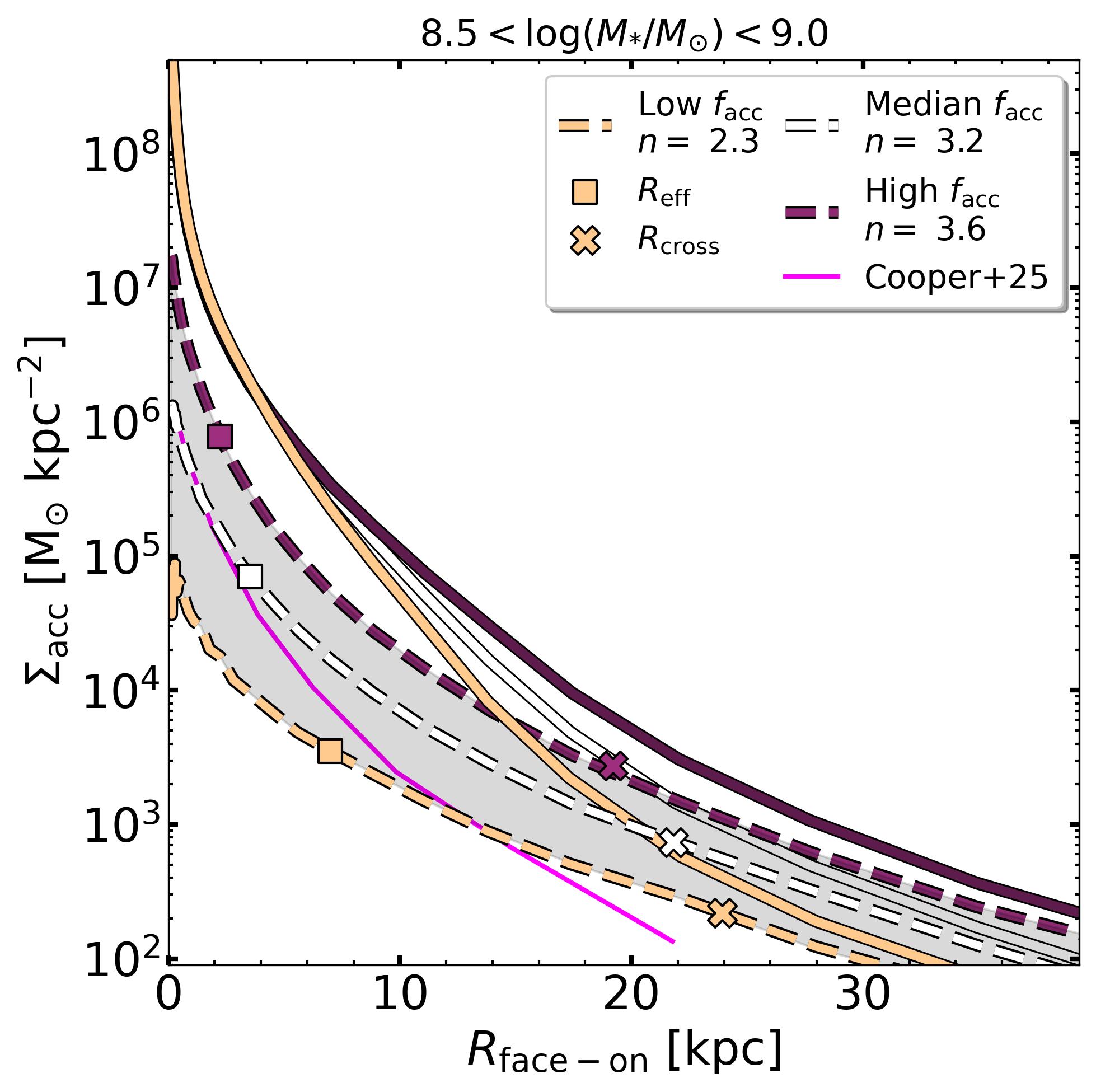}
    
    \caption{Surface density profiles of stacked galaxies of stellar mass in the range $8.5 < \log(M_{*})/M_\odot < 9.0$. Thick solid lines correspond to the surface stellar density profile of all stars, dashed lines for the accreted components, and in magenta we show the median profile reported by \citet{Cooper2025}. Three curves are shown for each: the white ones correspond top the median profile of all galaxies in that $M_*$ bin. Beige and purple curves correspond to those with lower-than-average and higher-than-average accreted mass fractions. More precisely, they correspond to those with $f_{\rm acc}/f_{\rm acc, med}<1/3$ and $f_{\rm acc}/f_{\rm acc, med}>3$, respectively. Open squares denote the effective radius of each set; crosses indicate the median value of the projected crossing radius, $R_{\rm cross}$, outside which the accreted component dominates over in-situ stars. S\'ersic-law fit indices, $n$, are also quoted in the legend. Note that the accreted profiles get more concentrated (i.e., lower $R_{\rm eff}$, larger $n$) with increasing $f_{\rm acc}$.}
    \label{fig:Profiles_Sérsic_fit_example}
\end{figure}

\begin{figure}
    \centering

    \includegraphics[width=\columnwidth]{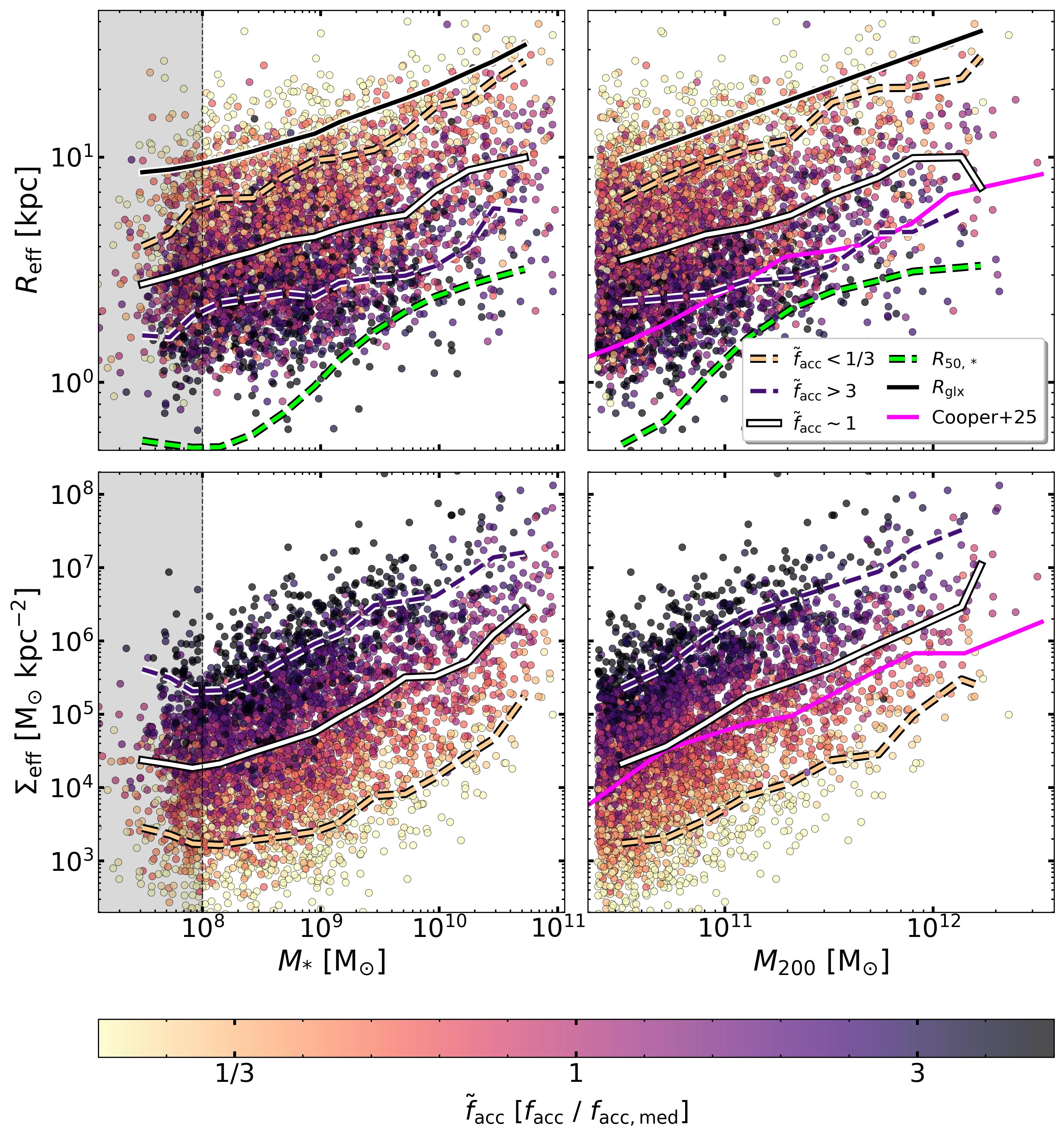}
    \includegraphics[width=\columnwidth]{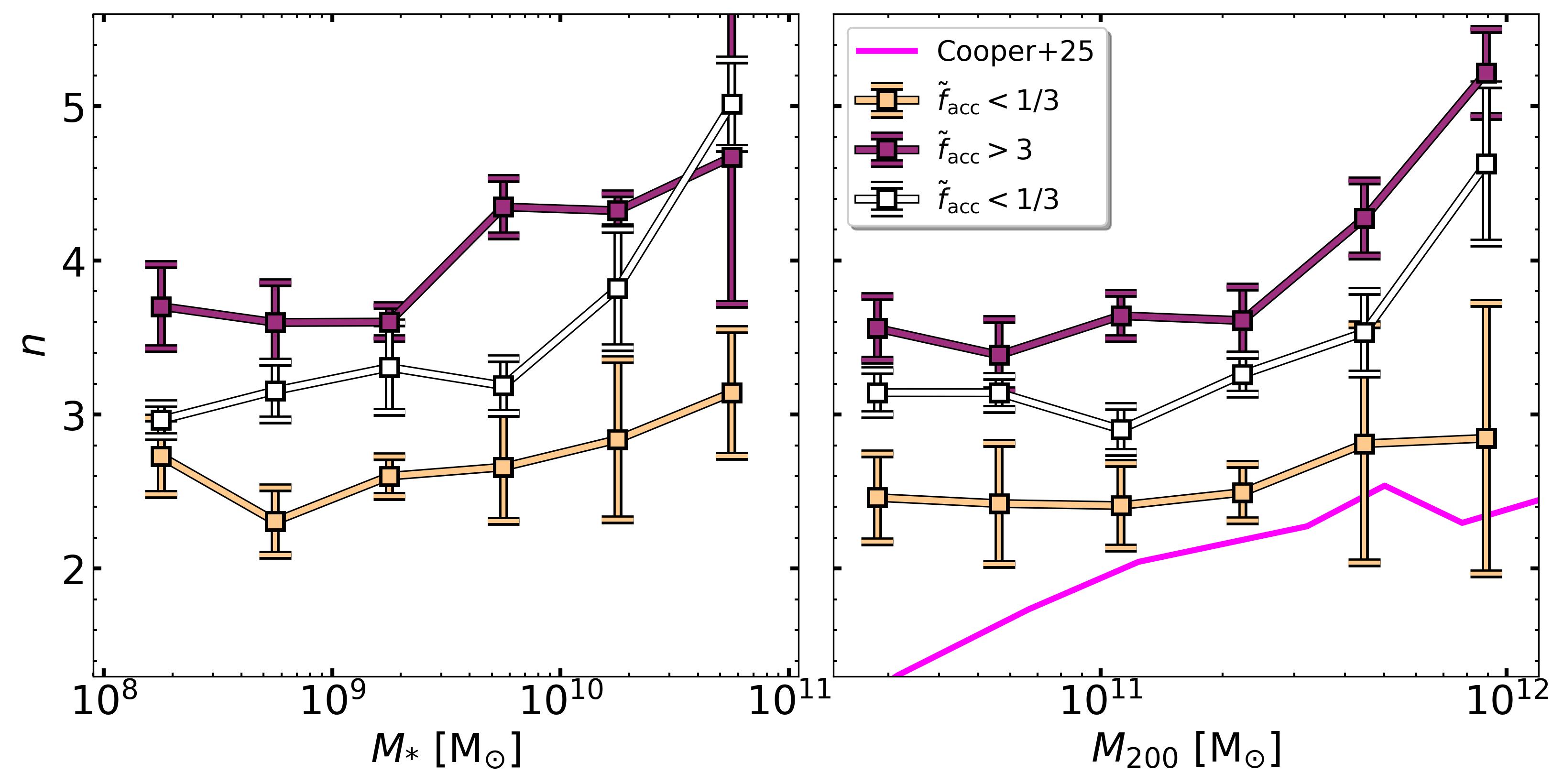}
    
    \caption{Effective radius $R_{\rm eff}$ (top panels) and surface density $\Sigma_{\rm eff}$ (centre panels) of the accreted component of all galaxies in our sample, coloured by accreted mass fraction, scaled to the median value at each mass (i.e., $\tilde{f}_{\rm acc}=f_{\rm acc}/f_{\rm acc, med}$), as a function of stellar mass (left panels) and virial mass (right panels). Dashed coloured lines show the median for galaxies with $\tilde{f}_{\rm acc}<1/3$ (beige) or $\tilde{f}_{\rm acc}>3$ (purple).  For comparison, in the top panels we show the median projected stellar half-mass radius of the total stellar component  (green dashed line) and the adopted galaxy radius  $R_{\rm glx}$ (in projection, black solid line). Sérsic indices $n$ derived from fits to the accreted component surface density profiles of stacked galaxies are shown in the bottom panels. With magenta solid lines we show the median trends reported by \citet{Cooper2025}.}
    \label{fig:Profiles_Sérsic_fit_props2}
\end{figure}

We study next the spatial distribution of accreted stars using  face-on surface density profiles stacked in narrow bins of stellar mass. An example is shown in Fig.~\ref{fig:Profiles_Sérsic_fit_example} for galaxies with $8.5<\log M_*/M_\odot<9.0$. Here  solid lines show the average stacked profile of all stars and dashed lines those of the accreted component. 

Three different profiles for each are shown: the median profiles are shown in white, whereas purple and beige  profiles correspond, respectively, to systems with unusually high or low accreted mass fractions. 
In particular, 'low $f_{\rm acc}$' refers to systems whose accreted fractions are below one-third of the median for that $M_*$ bin ($f_{\rm acc}/f_{\rm acc,med} = \tilde{f}_{\rm acc} <1/3$). Analogously, 'high $f_{\rm acc}$' denotes systems with $\tilde{f}_{\rm acc} > 3$.
Projected half-mass radii of the accreted component ($R_{\rm eff}$) are marked with filled squares and projected crossing radii ($R_{\rm cross}$) with filled crosses.

It is clear from Fig.~\ref{fig:Profiles_Sérsic_fit_example} that accreted mass profiles become more centrally concentrated as $f_{\rm acc}$ increases, which result in smaller values of $R_{\rm eff}$. As a result, the outermost portions of the accreted mass profile change less; indeed, at fixed radius the surface density of the 'high $f_{\rm acc}$' and  'low $f_{\rm acc}$' systems differs by a factor of $\sim$4 although the total accreted mass varies by more than one dex. This implies that even galaxies with rather different accreted components may have fairly similar outer stellar haloes.

We have fit these profiles using a S\'ersic law \citep[][]{Sersic1963} restricted to accreted particles with $R \gtrsim 2$ kpc to focus in the outer regions. S\'ersic's law,
\begin{equation}
    \Sigma(R) = \Sigma_\mathrm{eff}~\exp{\left\{ -b_n \left[ \left(\frac{R}{R_{\mathrm{eff}}} \right)^{1/n} -1 \right] \right\}},
\label{eq:Sersic_profile}
\end{equation}
has three fitting parameters, where $R_{\mathrm{eff}}$ and $\Sigma_\mathrm{eff}$ are the effective radius (containing half the stellar mass in projection) and the surface mass density at that radius, respectively. $n$ is the Sérsic index and $b_n$ follows from the definition of $R_{\mathrm{eff}}$: $b_n \approx 2n - 0.324$, for $n \gtrsim 1$. The $n$ value obtained for each accreted profile is indicated in the legend of the figure.

For comparison, the magenta curve shows the median density profile reported by \citet{Cooper2025} for the same bin of stellar mass. At inner regions, it matches the typical accreted component of TNG50 galaxies (grey shaded region), but exhibits a sharper decline, indicating a less extended accreted component than our results. Thus, our sample of galaxies exhibit typically higher accreted masses than those of \citet{Cooper2025}.

We use the fit parameters to illustrate how accreted mass profiles vary with galaxy and halo mass and also $\tilde{f}_{\rm acc}$ in Fig.~\ref{fig:Profiles_Sérsic_fit_props2}. There is clearly a large variation in the accreted profiles at given mass, but median trends still seem robust; The higher the accreted mass fraction (darker colours) the smaller the $R_{\rm eff}$ of the accreted component and, correspondingly, the higher $\Sigma_{\rm eff}$. 

This result is consistent with the idea that most accreted mass is contributed by just a few systems; the more massive the accreted system, the more resilient it is to tidal forces, and the closer to the centre it gets before getting disrupted by the host galaxy \citep[see e.g.][]{White1982, Dubinski1996, Naab2003}, leading to more concentrated accreted profiles. Indeed, haloes with the most accreted mass typically have density profiles with relatively small $R_{\rm eff}$, comparable to the main galaxy ($R_{50,*}$, green dashed line) and profile shapes as steep as $n \sim 4$, while those with the least accreted mass show $R_{\rm eff}$ as large as the fiducial galaxy size ($R_{\rm glx} = 0.25 ~ R_{200}$, black solid line) and  shallower profile shapes with $n \sim 2.5$.

Our results also compare favourably with those reported by \citet{Cooper2025}, despite their rather different numerical approach. The total accreted mass is typically higher in our analysis of TNG50, and, therefore, for similar values of $\Sigma_{\rm eff}$, the size of the accreted component at fixed mass is typically larger. The accreted component of TNG50 galaxies also extends up to larger galactocentric distances than those reported by \citet{Cooper2025}, exhibiting shallower density profiles at large radii, i.e. larger $n$, as shown in the bottom panels of Fig. \ref{fig:Profiles_Sérsic_fit_props2}.

\subsection{Stellar halo metallicities} \label{SecHaloMets}

\begin{figure}
    \centering
    \includegraphics[width=0.95\columnwidth]{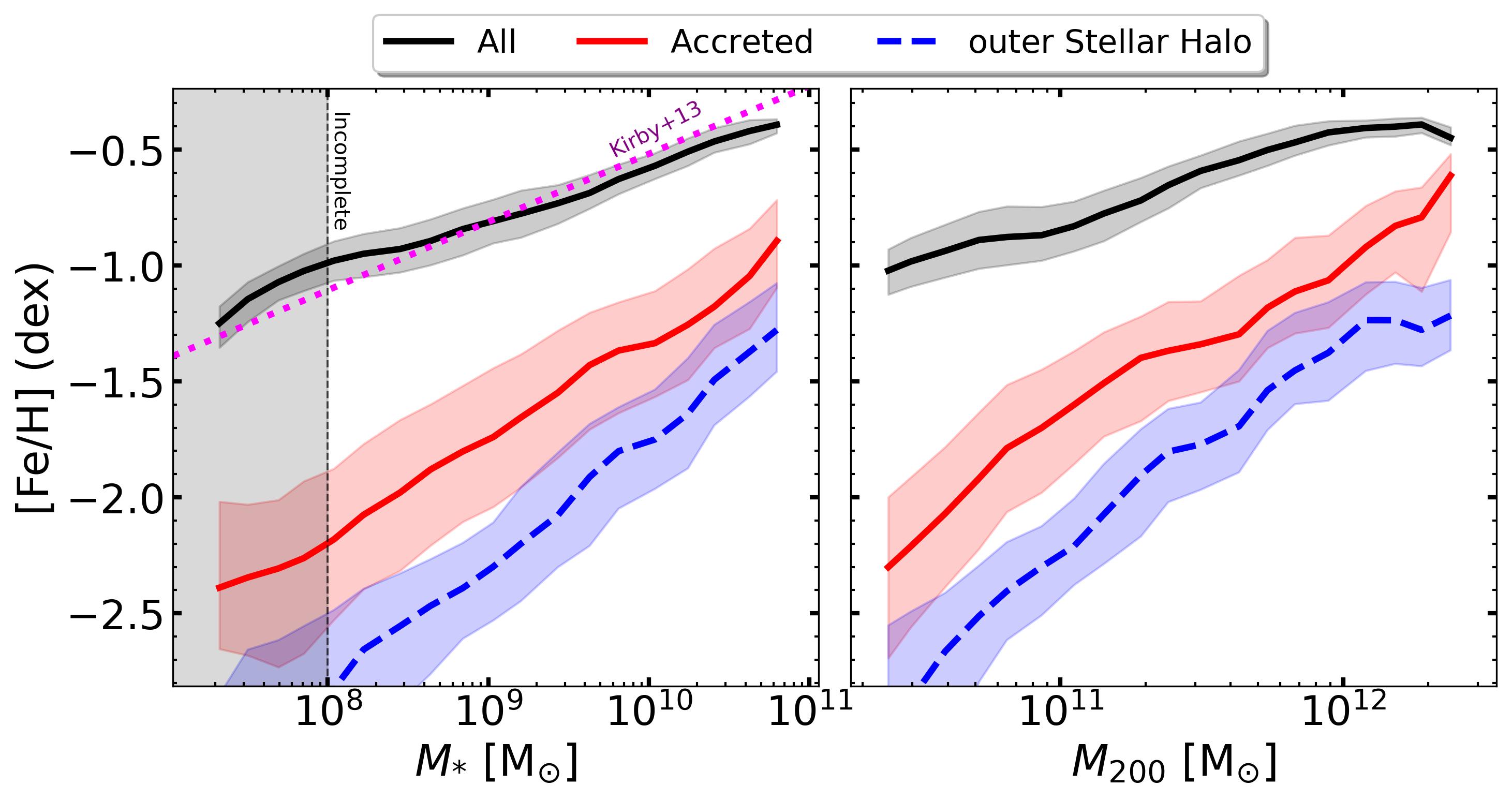}
    \caption{Stellar metallicity $\rm [Fe/H]$ for i) all stars bound to the main galaxy (solid black lines); ii) accreted stars (solid red lines); iii) and outer stellar halo (dashed blue lines), as a function of total stellar mass (left panel) or virial mass (right panel). The magenta dotted line  shows the mass-metallicity relation reported by \citet{Kirby2013}. Simulated metallicities have been shifted downwards by $-0.26$ dex to match the \citet{Kirby2013} relation at $M_*=10^9\, M_\odot$. Note the steep mass-metallicity of the accreted component, as well as the fact that outer stellar haloes have a substantial radial metallicity gradient. The outer haloes of dwarf galaxies should be fertile hunting grounds for metal-deficient stars.}
    \label{fig:t90_glx_acc_SH}
\end{figure}

We have seen above that in-situ stars are present throughout the halo, even in the outermost regions of a galaxy. How distinct are accreted stars from in-situ stars? A hint is provided by Fig.~\ref{fig:t90_glx_acc_SH}, which shows the median metallicity $\rm [Fe/H]$ (iron-to-hydrogen abundances, in solar units) of the various stellar components as a function of galaxy and halo mass. 

As shown by previous works \citep[see e.g.][]{Benavides2024, BianDu2025}, the median mass-metallicity trend of TNG50 galaxies has to be corrected to agree well with that of \citet[][magenta dotted line]{Kirby2013}. Here we show the TNG50 metallicities uniformly rescaled by $-0.26$ dex to match the \citet{Kirby2013} curve at $M_*=10^9\, M_\odot$.

Interestingly, the trend with mass of the accreted component metallicities is steeper that that of the main galaxies (red curves in Fig.~\ref{fig:t90_glx_acc_SH}), implying that accreted stars in dwarf galaxies are expected to be particularly poor in metal content. This trend is exacerbated when considering only the outer stellar halo (blue curves), mainly composed by accreted stars. In other words, the accreted stellar components have a substantial metallicity gradient. This is an intriguing result, as it suggests that some of the most metal poor stars in the Universe may be lurking in the far outskirts of dwarf galaxies \citep[see also][for a similar analysis]{Starkenburg2017}.

\section{Summary and Conclusions} \label{SecConc}

We have used the TNG50-1 cosmological hydrodynamic simulation to study the accreted stellar components of $\sim$5000 central galaxies spanning two orders of magnitude in virial mass ($10.3 \leq \log(M_{200}/M_{\odot}) \lesssim 12.3$) and roughly three decades in stellar mass ($8 \lesssim \log(M_{*}/M_{\odot}) \lesssim 11$). We define as 'accreted' stars formed physically far from the main progenitor of a galaxy ($r_{\mathrm{birth}} > r_{\mathrm{glx}}$), where the galaxy radius, $r_{\rm glx}$ is a fixed physical radius equal to $25\%$ of the virial radius of the system at $z=0$.

In agreement with previous  work, we find that the total accreted stellar mass correlates strongly with halo virial mass and galaxy mass. In TNG50, the median accreted mass fraction, $f_{\rm acc}$, decreases with decreasing galaxy mass from a peak of nearly $20\%$ at $M_*\sim$$10^{11}\, M_\odot$ down to $f_{\rm acc} < 0.02$ at $M_*\sim$$10^8\, M_\odot$, albeit with large scatter. Essentially all galaxies in the mass range probed by our sample have accreted stars, so stellar haloes are predicted to be a ubiquitous feature of galaxies in this mass range. 'Halo-free' galaxies, if they exist, are likely of masses not exceeding $M_* \sim 10^8\, M_\odot$.

The mass in the outer stellar halo, defined as stars found at present at radii beyond $r_{\rm glx}$, correlates strongly with the accreted mass, but it only contains $\sim$10\% of all accreted stars. Indeed, most accreted stars end up orbiting at small galactocentric distances rather than in the outer halo.

The galaxy radius, $r_{\rm glx}$, also roughly delineates the regions where the in-situ stellar component dominates ($r<r_{\rm glx}$) from the 'outer stellar halo' ($r>r_{\rm glx}$), where accreted stars dominate. This radius is roughly $7$-$10$ times larger than the half-mass radius of the main galaxy, implying that only studies of the far outskirts of a galaxy are likely to yield relatively pure samples of accreted stars.

At given mass, the spatial distribution of accreted stars depend strongly on the accreted mass fraction. The higher the accreted fraction, the more concentrated the accreted component becomes (i.e., smaller $R_{\rm eff}$ and larger values of S\'ersic index, $n$). The effective radius of the accreted component is typically larger than that of the main galaxy, and can exceed it by up to an order of magnitude. 

Our analysis generally confirms these earlier results \citep[see e.g.][for works at the Milky-Way scale]{Cooper2013,Pill2014,Amorisco2017}, although some quantitative differences are also found, some of which may be traced to differences in the definitions used to characterize accreted stars. For example, the  total accreted mass we report for TNG50 is higher (and the density profiles more concentrated) than those reported by \citet{Cooper2025}.

We confirm that accreted stars are typically more metal poor than the main galaxy, and that the metallicity of the accreted component scales with galaxy mass. There is also a strong metallicity gradient in the accreted component, so that accreted stars in the outer stellar halo are significantly more metal poor than the average accreted star. Our work extends these results to a statistically significant sample of field dwarf galaxies, and implies that the outer stellar haloes of dwarf galaxies should be a fertile hunting ground for extremely metal-poor stars.

These results also provide insight into the accreted mass and stellar halo relations for isolated galaxies across a wide range of stellar and virial masses within a cosmological context. This should help with the interpretation of upcoming data from observational projects such as LSST \citep{LSST2019}, Roman \citep{WFIRST2019} and Euclid \citep{Euclid2024}, which will resolve low surface brightness regions, such as the outskirts of bright nearby dwarfs. This insight will help to connect the faint stellar envelopes surrounding galaxies to their mass assembly history and accreted mass, two key components of the hierarchical structure formation predicted by the $\Lambda$CDM cosmology.

\begin{acknowledgements}
    We thank the referee Andrew Cooper for his insightful comments and a very constructive report of the manuscript. This project has received funding from the European Union’s HORIZON-MSCA-2021-SE-01 Research and Innovation programme under the Marie Sklodowska-Curie grant agreement number 101086388 - Project acronym: LACEGAL. This work was partially supported by the Consejo de Investigaciones Científicas y Técnicas de la República Argentina (CONICET) and the Secretaría de Ciencia y Técnica de la Universidad Nacional de Córdoba (SeCyT). The IllustrisTNG simulations were undertaken with compute time awarded by the Gauss Centre for Supercomputing (GCS) under GCS Large-Scale Projects GCS-ILLU and GCS-DWAR on the GCS share of the supercomputer Hazel Hen at the High Performance Computing Center Stuttgart (HLRS), as well as on the machines of the Max Planck Computing and Data Facility (MPCDF) in Garching, Germany. JFN acknowledges the hospitality of the Max-Planck Institute for Astrophysics,  the Donostia International Physics Center, and the Institute for Computational Cosmology at Durham University during the completion of this manuscript.      
\end{acknowledgements}

\bibliographystyle{aa}
\bibliography{aa56633-25}

\begin{thebibliography}{94}
\expandafter\ifx\csname natexlab\endcsname\relax\def\natexlab#1{#1}\fi

\bibitem[{{Abadi} {et~al.}(2006){Abadi}, {Navarro}, \& {Steinmetz}}]{Abadi2006}
{Abadi}, M.~G., {Navarro}, J.~F., \& {Steinmetz}, M. 2006, \mnras, 365, 747

\bibitem[{{Ahvazi} {et~al.}(2024){Ahvazi}, {Sales}, {Navarro}, {Benson}, {Boselli}, \& {D'Souza}}]{Ahvazi2024}
{Ahvazi}, N., {Sales}, L.~V., {Navarro}, J.~F., {et~al.} 2024, The Open Journal of Astrophysics, 7, 111

\bibitem[{{Akeson} {et~al.}(2019){Akeson}, {Armus}, {Bachelet}, {Bailey}, {Bartusek}, {Bellini}, {Benford}, {Bennett}, {Bhattacharya}, {Bohlin}, {Boyer}, {Bozza}, {Bryden}, {Calchi Novati}, {Carpenter}, {Casertano}, {Choi}, {Content}, {Dayal}, {Dressler}, {Dor{\'e}}, {Fall}, {Fan}, {Fang}, {Filippenko}, {Finkelstein}, {Foley}, {Furlanetto}, {Kalirai}, {Gaudi}, {Gilbert}, {Girard}, {Grady}, {Greene}, {Guhathakurta}, {Heinrich}, {Hemmati}, {Hendel}, {Henderson}, {Henning}, {Hirata}, {Ho}, {Huff}, {Hutter}, {Jansen}, {Jha}, {Johnson}, {Jones}, {Kasdin}, {Kelly}, {Kirshner}, {Koekemoer}, {Kruk}, {Lewis}, {Macintosh}, {Madau}, {Malhotra}, {Mandel}, {Massara}, {Masters}, {McEnery}, {McQuinn}, {Melchior}, {Melton}, {Mennesson}, {Peeples}, {Penny}, {Perlmutter}, {Pisani}, {Plazas}, {Poleski}, {Postman}, {Ranc}, {Rauscher}, {Rest}, {Roberge}, {Robertson}, {Rodney}, {Rhoads}, {Rhodes}, {Ryan}, {Sahu}, {Sand}, {Scolnic}, {Seth}, {Shvartzvald}, {Siellez}, {Smith}, {Spergel}, {Stassun}, {Street}, {Strolger}, {Szalay},
  {Trauger}, {Troxel}, {Turnbull}, {van der Marel}, {von der Linden}, {Wang}, {Weinberg}, {Williams}, {Windhorst}, {Wollack}, {Wu}, {Yee}, \& {Zimmerman}}]{WFIRST2019}
{Akeson}, R., {Armus}, L., {Bachelet}, E., {et~al.} 2019, arXiv e-prints, arXiv:1902.05569

\bibitem[{{Amorisco}(2017)}]{Amorisco2017}
{Amorisco}, N.~C. 2017, \mnras, 469, L48

\bibitem[{{Belokurov} {et~al.}(2018){Belokurov}, {Erkal}, {Evans}, {Koposov}, \& {Deason}}]{Belokurov2018}
{Belokurov}, V., {Erkal}, D., {Evans}, N.~W., {Koposov}, S.~E., \& {Deason}, A.~J. 2018, \mnras, 478, 611

\bibitem[{{Benavides} {et~al.}(2024){Benavides}, {Sales}, {Abadi}, {Vogelsberger}, {Marinacci}, \& {Hernquist}}]{Benavides2024}
{Benavides}, J.~A., {Sales}, L.~V., {Abadi}, M.~G., {et~al.} 2024, \apj, 977, 169

\bibitem[{{Ben{\'\i}tez-Llambay}(2017)}]{BenitezLlambay2017}
{Ben{\'\i}tez-Llambay}, A. 2017, {Py-SPHViewer: Cosmological simulations using Smoothed Particle Hydrodynamics}, Astrophysics Source Code Library, record ascl:1712.003

\bibitem[{{Ben{\'\i}tez-Llambay} {et~al.}(2016){Ben{\'\i}tez-Llambay}, {Navarro}, {Abadi}, {Gottl{\"o}ber}, {Yepes}, {Hoffman}, \& {Steinmetz}}]{B-L2016}
{Ben{\'\i}tez-Llambay}, A., {Navarro}, J.~F., {Abadi}, M.~G., {et~al.} 2016, \mnras, 456, 1185

\bibitem[{{Bian} {et~al.}(2025){Bian}, {Du}, {Debattista}, {Nelson}, {Norris}, {Ho}, {Lu}, {Cen}, {Ma}, {Ge}, {Fang}, \& {Li}}]{BianDu2025}
{Bian}, Y., {Du}, M., {Debattista}, V.~P., {et~al.} 2025, \apjl, 979, L33

\bibitem[{{Buder} {et~al.}(2025){Buder}, {Kos}, {Wang}, {McKenzie}, {Howell}, {Martell}, {Hayden}, {Zucker}, {Nordlander}, {Montet}, {Traven}, {Bland-Hawthorn}, {de Silva}, {Freeman}, {Lewis}, {Lind}, {Sharma}, {Simpson}, {Stello}, {Zwitter}, {Amarsi}, {Armstrong}, {Banks}, {Beavis}, {Beeson}, {Chen}, {Ciuc{\u{a}}}, {da Costa}, {de Grijs}, {Martin}, {Nataf}, {Ness}, {Rains}, {Scarr}, {Vogrin{\v{c}}i{\v{c}}}, {Wang}, {Wittenmyer}, {Xie}, \& {The Galah Collaboration}}]{GALAH2025}
{Buder}, S., {Kos}, J., {Wang}, X.~E., {et~al.} 2025, \pasa, 42, e051

\bibitem[{{Bullock} \& {Johnston}(2005)}]{BullockJohnston2005}
{Bullock}, J.~S. \& {Johnston}, K.~V. 2005, \apj, 635, 931

\bibitem[{{Carlin} {et~al.}(2019){Carlin}, {Garling}, {Peter}, {Crnojevi{\'c}}, {Forbes}, {Hargis}, {Mutlu-Pakdil}, {Pucha}, {Romanowsky}, {Sand}, {Spekkens}, {Strader}, \& {Willman}}]{Carlin2019}
{Carlin}, J.~L., {Garling}, C.~T., {Peter}, A. H.~G., {et~al.} 2019, \apj, 886, 109

\bibitem[{{Celiz} {et~al.}(2025){Celiz}, {Navarro}, {Abadi}, \& {Springel}}]{Celiz2025}
{Celiz}, B.~M., {Navarro}, J.~F., {Abadi}, M.~G., \& {Springel}, V. 2025, \aap, 699, A12

\bibitem[{{Chabrier}(2003)}]{Chabrier2003}
{Chabrier}, G. 2003, \pasp, 115, 763

\bibitem[{{Cooper} {et~al.}(2010){Cooper}, {Cole}, {Frenk}, {White}, {Helly}, {Benson}, {De Lucia}, {Helmi}, {Jenkins}, {Navarro}, {Springel}, \& {Wang}}]{Cooper2010}
{Cooper}, A.~P., {Cole}, S., {Frenk}, C.~S., {et~al.} 2010, \mnras, 406, 744

\bibitem[{{Cooper} {et~al.}(2013){Cooper}, {D'Souza}, {Kauffmann}, {Wang}, {Boylan-Kolchin}, {Guo}, {Frenk}, \& {White}}]{Cooper2013}
{Cooper}, A.~P., {D'Souza}, R., {Kauffmann}, G., {et~al.} 2013, \mnras, 434, 3348

\bibitem[{{Cooper} {et~al.}(2025){Cooper}, {Frenk}, {Hellwing}, \& {Bose}}]{Cooper2025}
{Cooper}, A.~P., {Frenk}, C.~S., {Hellwing}, W.~A., \& {Bose}, S. 2025, \mnras [\eprint[arXiv]{2501.13317}]

\bibitem[{{Cooper} {et~al.}(2015){Cooper}, {Parry}, {Lowing}, {Cole}, \& {Frenk}}]{Cooper2015}
{Cooper}, A.~P., {Parry}, O.~H., {Lowing}, B., {Cole}, S., \& {Frenk}, C. 2015, \mnras, 454, 3185

\bibitem[{{Davis} {et~al.}(1985){Davis}, {Efstathiou}, {Frenk}, \& {White}}]{Davies1985}
{Davis}, M., {Efstathiou}, G., {Frenk}, C.~S., \& {White}, S.~D.~M. 1985, \apj, 292, 371

\bibitem[{{De Almeida} {et~al.}(2024){De Almeida}, {Mamon}, {Dekel}, \& {Lima Neto}}]{Almeida2024}
{De Almeida}, A.~P., {Mamon}, G.~A., {Dekel}, A., \& {Lima Neto}, G.~B. 2024, \aap, 687, A131

\bibitem[{{De Lucia} \& {Helmi}(2008)}]{DeLuciaHelmi2008}
{De Lucia}, G. \& {Helmi}, A. 2008, \mnras, 391, 14

\bibitem[{{Deason} {et~al.}(2019){Deason}, {Belokurov}, \& {Sanders}}]{Deason2019}
{Deason}, A.~J., {Belokurov}, V., \& {Sanders}, J.~L. 2019, \mnras, 490, 3426

\bibitem[{{Deason} {et~al.}(2022){Deason}, {Bose}, {Fattahi}, {Amorisco}, {Hellwing}, \& {Frenk}}]{Deason2022}
{Deason}, A.~J., {Bose}, S., {Fattahi}, A., {et~al.} 2022, \mnras, 511, 4044

\bibitem[{{Dolag} {et~al.}(2009){Dolag}, {Borgani}, {Murante}, \& {Springel}}]{Dolag2009}
{Dolag}, K., {Borgani}, S., {Murante}, G., \& {Springel}, V. 2009, \mnras, 399, 497

\bibitem[{{D'Souza} \& {Bell}(2018)}]{DsouzaBell2018}
{D'Souza}, R. \& {Bell}, E.~F. 2018, \mnras, 474, 5300

\bibitem[{{D'Souza} {et~al.}(2014){D'Souza}, {Kauffman}, {Wang}, \& {Vegetti}}]{DSouza2014}
{D'Souza}, R., {Kauffman}, G., {Wang}, J., \& {Vegetti}, S. 2014, \mnras, 443, 1433

\bibitem[{{Dubinski} {et~al.}(1996){Dubinski}, {Mihos}, \& {Hernquist}}]{Dubinski1996}
{Dubinski}, J., {Mihos}, J.~C., \& {Hernquist}, L. 1996, \apj, 462, 576

\bibitem[{{Elias} {et~al.}(2018){Elias}, {Sales}, {Creasey}, {Cooper}, {Bullock}, {Rich}, \& {Hernquist}}]{Elias2018}
{Elias}, L.~M., {Sales}, L.~V., {Creasey}, P., {et~al.} 2018, \mnras, 479, 4004

\bibitem[{{Euclid Collaboration} {et~al.}(2024){Euclid Collaboration}, {Mellier}, {Abdurro'uf}, {Acevedo Barroso}, {Ach{\'u}carro}, {Adamek}, {Adam}, {Addison}, {Aghanim}, {Aguena}, {Ajani}, {Akrami}, {Al-Bahlawan}, {Alavi}, {Albuquerque}, {Alestas}, {Alguero}, {Allaoui}, {Allen}, {Allevato}, {Alonso-Tetilla}, {Altieri}, {Alvarez-Candal}, {Alvi}, {Amara}, {Amendola}, {Amiaux}, {Andika}, {Andreon}, {Andrews}, {Angora}, {Angulo}, {Annibali}, {Anselmi}, {Anselmi}, {Arcari}, {Archidiacono}, {Aric{\`o}}, {Arnaud}, {Arnouts}, {Asgari}, {Asorey}, {Atayde}, {Atek}, {Atrio-Barandela}, {Aubert}, {Aubourg}, {Auphan}, {Auricchio}, {Aussel}, {Aussel}, {Avelino}, {Avgoustidis}, {Avila}, {Awan}, {Azzollini}, {Baccigalupi}, {Bachelet}, {Bacon}, {Baes}, {Bagley}, {Bahr-Kalus}, {Balaguera-Antolinez}, {Balbinot}, {Balcells}, {Baldi}, {Baldry}, {Balestra}, {Ballardini}, {Ballester}, {Balogh}, {Ba{\~n}ados}, {Barbier}, {Bardelli}, {Baron}, {Barreiro}, {Barrena}, {Barriere}, {Barros}, {Barthelemy}, {Bartolo}, {Basset},
  {Battaglia}, {Battisti}, {Baugh}, {Baumont}, {Bazzanini}, {Beaulieu}, {Beckmann}, {Belikov}, {Bel}, {Bellagamba}, {Bella}, {Bellini}, {Benabed}, {Bender}, {Benevento}, {Bennett}, {Benson}, {Bergamini}, {Bermejo-Climent}, {Bernardeau}, {Bertacca}, {Berthe}, {Berthier}, {Bethermin}, {Beutler}, {Bevillon}, {Bhargava}, {Bhatawdekar}, {Bianchi}, {Bisigello}, {Biviano}, {Blake}, {Blanchard}, {Blazek}, {Blot}, {Bosco}, {Bodendorf}, {Boenke}, {B{\"o}hringer}, {Boldrini}, {Bolzonella}, {Bonchi}, {Bonici}, {Bonino}, {Bonino}, {Bonvin}, {Bon}, {Booth}, {Borgani}, {Borlaff}, {Borsato}, {Bosco}, {Bose}, {Botticella}, {Boucaud}, {Bouche}, {Boucher}, {Boutigny}, {Bouvard}, {Bouwens}, {Bouy}, {Bowler}, {Bozza}, {Bozzo}, {Branchini}, {Brando}, {Brau-Nogue}, {Brekke}, {Bremer}, {Brescia}, {Breton}, {Brinchmann}, {Brinckmann}, {Brockley-Blatt}, {Brodwin}, {Brouard}, {Brown}, {Bruton}, {Bucko}, {Buddelmeijer}, {Buenadicha}, {Buitrago}, {Burger}, {Burigana}, {Busillo}, {Busonero}, {Cabanac}, {Cabayol-Garcia}, {Cagliari},
  {Caillat}, {Caillat}, {Calabrese}, {Calabro}, {Calderone}, {Calura}, {Camacho Quevedo}, {Camera}, {Campos}, {Canas-Herrera}, {Candini}, {Cantiello}, {Capobianco}, {Cappellaro}, {Cappelluti}, {Cappi}, {Caputi}, {Cara}, {Carbone}, {Cardone}, {Carella}, {Carlberg}, {Carle}, {Carminati}, {Caro}, {Carrasco}, {Carretero}, {Carrilho}, {Carron Duque}, {Carry}, {Carvalho}, {Carvalho}, {Casas}, {Casas}, {Casenove}, {Casey}, {Cassata}, {Castander}, {Castelao}, {Castellano}, {Castiblanco}, {Castignani}, {Castro}, {Cavet}, {Cavuoti}, {Chabaud}, {Chambers}, {Charles}, {Charlot}, {Chartab}, {Chary}, {Chaumeil}, {Cho}, {Chon}, {Ciancetta}, {Ciliegi}, {Cimatti}, {Cimino}, {Cioni}, {Claydon}, {Cleland}, {Cl{\'e}ment}, {Clements}, {Clerc}, {Clesse}, {Codis}, {Cogato}, {Colbert}, {Cole}, {Coles}, {Collett}, {Collins}, {Colodro-Conde}, {Colombo}, {Combes}, {Conforti}, {Congedo}, {Conseil}, {Conselice}, {Contarini}, {Contini}, {Conversi}, {Cooray}, {Copin}, {Corasaniti}, {Corcho-Caballero}, {Corcione}, {Cordes}, {Corpace},
  {Correnti}, {Costanzi}, {Costille}, {Courbin}, {Courcoult Mifsud}, {Courtois}, {Cousinou}, {Covone}, {Cowell}, {Cragg}, {Cresci}, {Cristiani}, {Crocce}, {Cropper}, {E Crouzet}, {Csizi}, {Cuby}, {Cucchetti}, {Cucciati}, {Cuillandre}, {Cunha}, {Cuozzo}, {Daddi}, {D'Addona}, {Dafonte}, {Dagoneau}, {Dalessandro}, {Dalton}, {D'Amico}, {Dannerbauer}, {Danto}, {Das}, {Da Silva}, {da Silva}, {d'Assignies Doumerg}, {Daste}, {Davies}, {Davini}, {Dayal}, {de Boer}, {Decarli}, {De Caro}, {Degaudenzi}, {Degni}, {de Jong}, {de la Bella}, {de la Torre}, {Delhaise}, {Delley}, {Delucchi}, {De Lucia}, {Denniston}, {De Paolis}, {De Petris}, {Derosa}, {Desai}, {Desjacques}, {Despali}, {Desprez}, {De Vicente-Albendea}, {Deville}, {Dias}, {D{\'\i}az-S{\'a}nchez}, {Diaz}, {Di Domizio}, {Diego}, {Di Ferdinando}, {Di Giorgio}, {Dimauro}, {Dinis}, {Dolag}, {Dolding}, {Dole}, {Dom{\'\i}nguez S{\'a}nchez}, {Dor{\'e}}, {Dournac}, {Douspis}, {Dreihahn}, {Droge}, {Dryer}, {Dubath}, {Duc}, {Ducret}, {Duffy}, {Dufresne}, {Duncan}, {Dupac},
  {Duret}, {Durrer}, {Durret}, {Dusini}, {Ealet}, {Eggemeier}, {Eisenhardt}, {Elbaz}, {Elkhashab}, {Ellien}, {Endicott}, {Enia}, {Erben}, {Escartin Vigo}, {Escoffier}, {Escudero Sanz}, {Essert}, {Ettori}, {Ezziati}, {Fabbian}, {Fabricius}, {Fang}, {Farina}, {Farina}, {Farinelli}, {Farrens}, {Faustini}, {Feltre}, {Ferguson}, {Ferrando}, {Ferrari}, {Ferr{\'e}-Mateu}, {Ferreira}, {Ferreras}, {Ferrero}, {Ferriol}, {Ferruit}, {Filleul}, {Finelli}, {Finkelstein}, {Finoguenov}, {Fiorini}, {Flentge}, {Focardi}, {Fonseca}, {Fontana}, {Fontanot}, {Fornari}, {Fosalba}, {Fossati}, {Fotopoulou}, {Fouchez}, {Fourmanoit}, {Frailis}, {Fraix-Burnet}, {Franceschi}, {Franco}, {Franzetti}, {Freihoefer}, {Frenk}, {Frittoli}, {Frugier}, {Frusciante}, {Fumagalli}, {Fumagalli}, {Fumana}, {Fu}, {Gabarra}, {Galeotta}, {Galluccio}, {Ganga}, {Gao}, {Garc{\'\i}a-Bellido}, {Garcia}, {Gardner}, {Garilli}, {Gaspar-Venancio}, {Gasparetto}, {Gautard}, {Gavazzi}, {Gaztanaga}, {Genolet}, {Genova Santos}, {Gentile}, {George}, {Gerbino},
  {Ghaffari}, {Giacomini}, {Gianotti}, {Gibb}, {Gillard}, {Gillis}, {Ginolfi}, {Giocoli}, {Girardi}, {Giri}, {Goh}, {G{\'o}mez-Alvarez}, {Gonzalez-Perez}, {Gonzalez}, {Gonzalez}, {Gonzalez}, {Gouyou Beauchamps}, {Gozaliasl}, {Gracia-Carpio}, {Grandis}, {Granett}, {Granvik}, {Grazian}, {Gregorio}, {Grenet}, {Grillo}, {Grupp}, {Gruppioni}, {Gruppuso}, {Guerbuez}, {Guerrini}, {Guidi}, {Guillard}, {Gutierrez}, {Guttridge}, {Guzzo}, {Gwyn}, {Haapala}, {Haase}, {Haddow}, {Hailey}, {Hall}, {Hall}, {Hamaus}, {Haridasu}, {Harnois-D{\'e}raps}, {Harper}, {Hartley}, {Hasinger}, {Hassani}, {Hatch}, {Haugan}, {H{\"a}u{\ss}ler}, {Heavens}, {Heisenberg}, {Helmi}, {Helou}, {Hemmati}, {Henares}, {Herent}, {Hern{\'a}ndez-Monteagudo}, {Heuberger}, {Hewett}, {Heydenreich}, {Hildebrandt}, {Hirschmann}, {Hjorth}, {Hoar}, {Hoekstra}, {Holland}, {Holliman}, {Holmes}, {Hook}, {Horeau}, {Hormuth}, {Hornstrup}, {Hosseini}, {Hu}, {Hudelot}, {Hudson}, {Huertas-Company}, {Huff}, {Hughes}, {Humphrey}, {Hunt}, {Huynh}, {Ibata}, {Ichikawa},
  {Iglesias-Groth}, {Ilbert}, {Ili{\'c}}, {Ingoglia}, {Iodice}, {Israel}, {Israelsson}, {Izzo}, {Jablonka}, {Jackson}, {Jacobson}, {Jafariyazani}, {Jahnke}, {Jain}, {Jansen}, {Jarvis}, {Jasche}, {Jauzac}, {Jeffrey}, {Jhabvala}, {Jimenez-Teja}, {Jimenez Mu{\~n}oz}, {Joachimi}, {Johansson}, {Joudaki}, {Jullo}, {Kajava}, {Kang}, {Kannawadi}, {Kansal}, {Karagiannis}, {K{\"a}rcher}, {Kashlinsky}, {Kazandjian}, {Keck}, {Keih{\"a}nen}, {Kerins}, {Kermiche}, {Khalil}, {Kiessling}, {Kiiveri}, {Kilbinger}, {Kim}, {King}, {Kirkpatrick}, {Kitching}, {Kluge}, {Knabenhans}, {Knapen}, {Knebe}, {Kneib}, {Kohley}, {Koopmans}, {Koskinen}, {Koulouridis}, {Kou}, {Kov{\'a}cs}, {Kova{\v{c}}i{\'c}}, {Kowalczyk}, {Koyama}, {Kraljic}, {Krause}, {Kruk}, {Kubik}, {Kuchner}, {Kuijken}, {K{\"u}mmel}, {Kunz}, {Kurki-Suonio}, {Lacasa}, {Lacey}, {La Franca}, {Lagarde}, {Lahav}, {Laigle}, {La Marca}, {La Marle}, {Lamine}, {Lam}, {Lan{\c{c}}on}, {Landt}, {Langer}, {Lapi}, {Larcheveque}, {Larsen}, {Lattanzi}, {Laudisio}, {Laugier}, {Laureijs},
  {Laurent}, {Lavaux}, {Lawrenson}, {Lazanu}, {Lazeyras}, {Le Boulc'h}, {Le Brun}, {Le Brun}, {Leclercq}, {Lee}, {Le Graet}, {Legrand}, {Leirvik}, {Le Jeune}, {Lembo}, {Le Mignant}, {Lepinzan}, {Lepori}, {Le Reun}, {Leroy}, {Lesci}, {Lesgourgues}, {Leuzzi}, {Levi}, {Liaudat}, {Libet}, {Liebing}, {Ligori}, {Lilje}, {Lin}, {Linde}, {Linder}, {Lindholm}, {Linke}, {Li}, {Liu}, {Lloro}, {Lobo}, {Lodieu}, {Lombardi}, {Lombriser}, {Lonare}, {Longo}, {L{\'o}pez-Caniego}, {Lopez Lopez}, {Alvarez}, {Loureiro}, {Loveday}, {Lusso}, {Macias-Perez}, {Maciaszek}, {Maggio}, {Magliocchetti}, {Magnard}, {Magnier}, {Magro}, {Mahler}, {Mainetti}, {Maino}, {Maiorano}, {Maiorano}, {Malavasi}, {Mamon}, {Mancini}, {Mandelbaum}, {Manera}, {Manj{\'o}n-Garc{\'\i}a}, {Mannucci}, {Mansutti}, {Manteiga Outeiro}, {Maoli}, {Maraston}, {Marcin}, {Marcos-Arenal}, {Margalef-Bentabol}, {Marggraf}, {Marinucci}, {Marinucci}, {Markovic}, {Marleau}, {Marpaud}, {Martignac}, {Mart{\'\i}n-Fleitas}, {Martin-Moruno}, {Martin}, {Martinelli}, {Martinet},
  {Martin}, {Martins}, {Marulli}, {Massari}, {Massey}, {Masters}, {Matarrese}, {Matsuoka}, {Matthew}, {Maughan}, {Mauri}, {Maurin}, {Maurogordato}, {McCarthy}, {McConnachie}, {McCracken}, {McDonald}, {McEwen}, {McPartland}, {Medinaceli}, {Mehta}, {Mei}, {Melchior}, {Melin}, {M{\'e}nard}, {Mendes}, {Mendez-Abreu}, {Meneghetti}, {Mercurio}, {Merlin}, {Metcalf}, {Meylan}, {Migliaccio}, {Mignoli}, {Miller}, {Miluzio}, {Milvang-Jensen}, {Mimoso}, {Miquel}, {Miyatake}, {Mobasher}, {Mohr}, {Monaco}, {Mongui{\'o}}, {Montoro}, {Mora}, {Moradinezhad Dizgah}, {Moresco}, {Moretti}, {Morgante}, {Morisset}, {Moriya}, {Morris}, {Mortlock}, {Moscardini}, {Mota}, {Mottet}, {Moustakas}, {Moutard}, {M{\"u}ller}, {Munari}, {Murphree}, {Murray}, {Murray}, {Musi}, {Nadathur}, {Nagam}, {Nagao}, {Naidoo}, {Nakajima}, {Nally}, {Natoli}, {Navarro-Alsina}, {Navarro Girones}, {Neissner}, {Nersesian}, {Nesseris}, {Nguyen-Kim}, {Nicastro}, {Nichol}, {Nielbock}, {Niemi}, {Nieto}, {Nilsson}, {Noller}, {Norberg}, {Nouri-Zonoz}, {Ntelis},
  {Nucita}, {Nugent}, {Nunes}, {Nutma}, {Ocampo}, {Odier}, {Oesch}, {Oguri}, {Magalhaes Oliveira}, {Onoue}, {Oosterbroek}, {Oppizzi}, {Ordenovic}, {Osato}, {Pacaud}, {Pace}, {Padilla}, {Paech}, {Pagano}, {Page}, {Palazzi}, {Paltani}, {Pamuk}, {Pandolfi}, {Paoletti}, {Paolillo}, {Papaderos}, {Pardede}, {Parimbelli}, {Parmar}, {Partmann}, {Pasian}, {Passalacqua}, {Paterson}, {Patrizii}, {Pattison}, {Paulino-Afonso}, {Paviot}, {Peacock}, {Pearce}, {Pedersen}, {Peel}, {Peletier}, {Pellejero Ibanez}, {Pello}, {Penny}, {Percival}, {Perez-Garrido}, {Perotto}, {Pettorino}, {Pezzotta}, {Pezzuto}, {Philippon}, {Pierre}, {Piersanti}, {Pietroni}, {Piga}, {Pilo}, {Pires}, {Pisani}, {Pizzella}, {Pizzuti}, {Plana}, {Polenta}, {Pollack}, {Poncet}, {P{\"o}ntinen}, {Pool}, {Popa}, {Popa}, {Popp}, {Porciani}, {Porth}, {Potter}, {Poulain}, {Pourtsidou}, {Pozzetti}, {Prandoni}, {Pratt}, {Prezelus}, {Prieto}, {Pugno}, {Quai}, {Quilley}, {Racca}, {Raccanelli}, {R{\'a}cz}, {Radinovi{\'c}}, {Radovich}, {Ragagnin}, {Ragnit}, {Raison},
  {Ramos-Chernenko}, {Ranc}, {Rasera}, {Raylet}, {Rebolo}, {Refregier}, {Reimberg}, {Reiprich}, {Renk}, {Renzi}, {Retre}, {Revaz}, {Reyl{\'e}}, {Reynolds}, {Rhodes}, {Ricci}, {Ricci}, {Riccio}, {Ricken}, {Rissanen}, {Risso}, {Rix}, {Robin}, {Rocca-Volmerange}, {Rocci}, {Rodenhuis}, {Rodighiero}, {Rodriguez Monroy}, {Rollins}, {Romanello}, {Roman}, {Romelli}, {Romero-Gomez}, {Roncarelli}, {Rosati}, {Rosset}, {Rossetti}, {Roster}, {Rottgering}, {Rozas-Fern{\'a}ndez}, {Ruane}, {Rubino-Martin}, {Rudolph}, {Ruppin}, {Rusholme}, {Sacquegna}, {S{\'a}ez-Casares}, {Saga}, {Saglia}, {Sahl{\'e}n}, {Saifollahi}, {Sakr}, {Salvalaggio}, {Salvaterra}, {Salvati}, {Salvato}, {Salvignol}, {S{\'a}nchez}, {Sanchez}, {Sanders}, {Sapone}, {Saponara}, {Sarpa}, {Sarron}, {Sartori}, {Sartoris}, {Sassolas}, {Sauniere}, {Sauvage}, {Sawicki}, {Scaramella}, {Scarlata}, {Scharr{\'e}}, {Schaye}, {Schewtschenko}, {Schindler}, {Schinnerer}, {Schirmer}, {Schmidt}, {Schmidt}, {Schmidt}, {Schneider}, {Schneider}, {Schneider}, {Sch{\"o}neberg},
  {Schrabback}, {Schultheis}, {Schulz}, {Schuster}, {Schwartz}, {Sciotti}, {Scodeggio}, {Scognamiglio}, {Scott}, {Scottez}, {Secroun}, {Sefusatti}, {Seidel}, {Seiffert}, {Sellentin}, {Selwood}, {Semboloni}, {Sereno}, {Serjeant}, {Serrano}, {Setnikar}, {Shankar}, {Sharples}, {Short}, {Shulevski}, {Shuntov}, {Sias}, {Sikkema}, {Silvestri}, {Simon}, {Sirignano}, {Sirri}, {Skottfelt}, {Slezak}, {Sluse}, {Smith}, {Smith}, {Smith}, {Smit}, {Soldano}, {Solheim}, {Sorce}, {Sorrenti}, {Soubrie}, {Spinoglio}, {Spurio Mancini}, {Stadel}, {Stagnaro}, {Stanco}, {Stanford}, {Starck}, {Stassi}, {Steinwagner}, {Stern}, {Stone}, {Strada}, {Strafella}, {Stramaccioni}, {Surace}, {Sureau}, {Suyu}, {Swindells}, {Szafraniec}, {Szapudi}, {Taamoli}, {Talia}, {Tallada-Cresp{\'\i}}, {Tanidis}, {Tao}, {Tarr{\'\i}o}, {Tavagnacco}, {Taylor}, {Taylor}, {Taylor}, {Teixeira}, {Tenti}, {Teodoro Idiago}, {Teplitz}, {Tereno}, {Tessore}, {Testa}, {Testera}, {Tewes}, {Teyssier}, {Theret}, {Thizy}, {Thomas}, {Toba}, {Toft}, {Toledo-Moreo},
  {Tolstoy}, {Tommasi}, {Torbaniuk}, {Torradeflot}, {Tortora}, {Tosi}, {Tosti}, {Trifoglio}, {Troja}, {Trombetti}, {Tronconi}, {Tsedrik}, {Tsyganov}, {Tucci}, {Tutusaus}, {Uhlemann}, {Ulivi}, {Urbano}, {Vacher}, {Vaillon}, {Valageas}, {Valdes}, {Valentijn}, {Valenziano}, {Valieri}, {Valiviita}, {Van den Broeck}, {Vassallo}, {Vavrek}, {Vega-Ferrero}, {Venemans}, {Venhola}, {Ventura}, {Verdoes Kleijn}, {Vergani}, {Verma}, {Vernizzi}, {Veropalumbo}, {Verza}, {Vescovi}, {Vibert}, {Viel}, {Vielzeuf}, {Viglione}, {Viitanen}, {Villaescusa-Navarro}, {Vinciguerra}, {Visticot}, {Voggel}, {von Wietersheim-Kramsta}, {Vriend}, {Wachter}, {Walmsley}, {Walth}, {Walton}, {Walton}, {Wander}, {Wang}, {Wang}, {Weaver}, {Weller}, {Wetzstein}, {Whalen}, {Whittam}, {Widmer}, {Wiesmann}, {Wilde}, {Williams}, {Winther}, {Wittje}, {Wong}, {Wright}, {Yankelevich}, {Yeung}, {Yoon}, {Youles}, {Yung}, {Zacchei}, {Zalesky}, {Zamorani}, {Zamorano Vitorelli}, {Zanoni Marc}, {Zennaro}, {Zerbi}, {Zinchenko}, {Zoubian}, {Zucca}, \&
  {Zumalacarregui}}]{Euclid2024}
{Euclid Collaboration}, {Mellier}, Y., {Abdurro'uf}, {et~al.} 2024, arXiv e-prints, arXiv:2405.13491

\bibitem[{{Fielder} {et~al.}(2024){Fielder}, {Jones}, {Sand}, {Bennet}, {Crnojevi{\'c}}, {Karunakaran}, {Mutlu-Pakdil}, \& {Spekkens}}]{Fielder2024}
{Fielder}, C., {Jones}, M.~G., {Sand}, D.~J., {et~al.} 2024, \aj, 168, 212

\bibitem[{{Freeman} \& {Bland-Hawthorn}(2002)}]{FreemanBH2002}
{Freeman}, K. \& {Bland-Hawthorn}, J. 2002, \araa, 40, 487

\bibitem[{{Frenk} {et~al.}(1988){Frenk}, {White}, {Davis}, \& {Efstathiou}}]{Frenk1988}
{Frenk}, C.~S., {White}, S. D.~M., {Davis}, M., \& {Efstathiou}, G. 1988, \apj, 327, 507

\bibitem[{{Gaia Collaboration} {et~al.}(2018){Gaia Collaboration}, {Brown}, {Vallenari}, {Prusti}, {de Bruijne}, {Babusiaux}, {Bailer-Jones}, {Biermann}, {Evans}, {Eyer}, {Jansen}, {Jordi}, {Klioner}, {Lammers}, {Lindegren}, {Luri}, {Mignard}, {Panem}, {Pourbaix}, {Randich}, {Sartoretti}, {Siddiqui}, {Soubiran}, {van Leeuwen}, {Walton}, {Arenou}, {Bastian}, {Cropper}, {Drimmel}, {Katz}, {Lattanzi}, {Bakker}, {Cacciari}, {Casta{\~n}eda}, {Chaoul}, {Cheek}, {De Angeli}, {Fabricius}, {Guerra}, {Holl}, {Masana}, {Messineo}, {Mowlavi}, {Nienartowicz}, {Panuzzo}, {Portell}, {Riello}, {Seabroke}, {Tanga}, {Th{\'e}venin}, {Gracia-Abril}, {Comoretto}, {Garcia-Reinaldos}, {Teyssier}, {Altmann}, {Andrae}, {Audard}, {Bellas-Velidis}, {Benson}, {Berthier}, {Blomme}, {Burgess}, {Busso}, {Carry}, {Cellino}, {Clementini}, {Clotet}, {Creevey}, {Davidson}, {De Ridder}, {Delchambre}, {Dell'Oro}, {Ducourant}, {Fern{\'a}ndez-Hern{\'a}ndez}, {Fouesneau}, {Fr{\'e}mat}, {Galluccio}, {Garc{\'\i}a-Torres},
  {Gonz{\'a}lez-N{\'u}{\~n}ez}, {Gonz{\'a}lez-Vidal}, {Gosset}, {Guy}, {Halbwachs}, {Hambly}, {Harrison}, {Hern{\'a}ndez}, {Hestroffer}, {Hodgkin}, {Hutton}, {Jasniewicz}, {Jean-Antoine-Piccolo}, {Jordan}, {Korn}, {Krone-Martins}, {Lanzafame}, {Lebzelter}, {L{\"o}ffler}, {Manteiga}, {Marrese}, {Mart{\'\i}n-Fleitas}, {Moitinho}, {Mora}, {Muinonen}, {Osinde}, {Pancino}, {Pauwels}, {Petit}, {Recio-Blanco}, {Richards}, {Rimoldini}, {Robin}, {Sarro}, {Siopis}, {Smith}, {Sozzetti}, {S{\"u}veges}, {Torra}, {van Reeven}, {Abbas}, {Abreu Aramburu}, {Accart}, {Aerts}, {Altavilla}, {{\'A}lvarez}, {Alvarez}, {Alves}, {Anderson}, {Andrei}, {Anglada Varela}, {Antiche}, {Antoja}, {Arcay}, {Astraatmadja}, {Bach}, {Baker}, {Balaguer-N{\'u}{\~n}ez}, {Balm}, {Barache}, {Barata}, {Barbato}, {Barblan}, {Barklem}, {Barrado}, {Barros}, {Barstow}, {Bartholom{\'e} Mu{\~n}oz}, {Bassilana}, {Becciani}, {Bellazzini}, {Berihuete}, {Bertone}, {Bianchi}, {Bienaym{\'e}}, {Blanco-Cuaresma}, {Boch}, {Boeche}, {Bombrun}, {Borrachero},
  {Bossini}, {Bouquillon}, {Bourda}, {Bragaglia}, {Bramante}, {Breddels}, {Bressan}, {Brouillet}, {Br{\"u}semeister}, {Brugaletta}, {Bucciarelli}, {Burlacu}, {Busonero}, {Butkevich}, {Buzzi}, {Caffau}, {Cancelliere}, {Cannizzaro}, {Cantat-Gaudin}, {Carballo}, {Carlucci}, {Carrasco}, {Casamiquela}, {Castellani}, {Castro-Ginard}, {Charlot}, {Chemin}, {Chiavassa}, {Cocozza}, {Costigan}, {Cowell}, {Crifo}, {Crosta}, {Crowley}, {Cuypers}, {Dafonte}, {Damerdji}, {Dapergolas}, {David}, {David}, {de Laverny}, {De Luise}, {De March}, {de Martino}, {de Souza}, {de Torres}, {Debosscher}, {del Pozo}, {Delbo}, {Delgado}, {Delgado}, {Di Matteo}, {Diakite}, {Diener}, {Distefano}, {Dolding}, {Drazinos}, {Dur{\'a}n}, {Edvardsson}, {Enke}, {Eriksson}, {Esquej}, {Eynard Bontemps}, {Fabre}, {Fabrizio}, {Faigler}, {Falc{\~a}o}, {Farr{\`a}s Casas}, {Federici}, {Fedorets}, {Fernique}, {Figueras}, {Filippi}, {Findeisen}, {Fonti}, {Fraile}, {Fraser}, {Fr{\'e}zouls}, {Gai}, {Galleti}, {Garabato}, {Garc{\'\i}a-Sedano}, {Garofalo},
  {Garralda}, {Gavel}, {Gavras}, {Gerssen}, {Geyer}, {Giacobbe}, {Gilmore}, {Girona}, {Giuffrida}, {Glass}, {Gomes}, {Granvik}, {Gueguen}, {Guerrier}, {Guiraud}, {Guti{\'e}rrez-S{\'a}nchez}, {Haigron}, {Hatzidimitriou}, {Hauser}, {Haywood}, {Heiter}, {Helmi}, {Heu}, {Hilger}, {Hobbs}, {Hofmann}, {Holland}, {Huckle}, {Hypki}, {Icardi}, {Jan{\ss}en}, {Jevardat de Fombelle}, {Jonker}, {Juh{\'a}sz}, {Julbe}, {Karampelas}, {Kewley}, {Klar}, {Kochoska}, {Kohley}, {Kolenberg}, {Kontizas}, {Kontizas}, {Koposov}, {Kordopatis}, {Kostrzewa-Rutkowska}, {Koubsky}, {Lambert}, {Lanza}, {Lasne}, {Lavigne}, {Le Fustec}, {Le Poncin-Lafitte}, {Lebreton}, {Leccia}, {Leclerc}, {Lecoeur-Taibi}, {Lenhardt}, {Leroux}, {Liao}, {Licata}, {Lindstr{\o}m}, {Lister}, {Livanou}, {Lobel}, {L{\'o}pez}, {Managau}, {Mann}, {Mantelet}, {Marchal}, {Marchant}, {Marconi}, {Marinoni}, {Marschalk{\'o}}, {Marshall}, {Martino}, {Marton}, {Mary}, {Massari}, {Matijevi{\v{c}}}, {Mazeh}, {McMillan}, {Messina}, {Michalik}, {Millar}, {Molina}, {Molinaro},
  {Moln{\'a}r}, {Montegriffo}, {Mor}, {Morbidelli}, {Morel}, {Morris}, {Mulone}, {Muraveva}, {Musella}, {Nelemans}, {Nicastro}, {Noval}, {O'Mullane}, {Ord{\'e}novic}, {Ord{\'o}{\~n}ez-Blanco}, {Osborne}, {Pagani}, {Pagano}, {Pailler}, {Palacin}, {Palaversa}, {Panahi}, {Pawlak}, {Piersimoni}, {Pineau}, {Plachy}, {Plum}, {Poggio}, {Poujoulet}, {Pr{\v{s}}a}, {Pulone}, {Racero}, {Ragaini}, {Rambaux}, {Ramos-Lerate}, {Regibo}, {Reyl{\'e}}, {Riclet}, {Ripepi}, {Riva}, {Rivard}, {Rixon}, {Roegiers}, {Roelens}, {Romero-G{\'o}mez}, {Rowell}, {Royer}, {Ruiz-Dern}, {Sadowski}, {Sagrist{\`a} Sell{\'e}s}, {Sahlmann}, {Salgado}, {Salguero}, {Sanna}, {Santana-Ros}, {Sarasso}, {Savietto}, {Schultheis}, {Sciacca}, {Segol}, {Segovia}, {S{\'e}gransan}, {Shih}, {Siltala}, {Silva}, {Smart}, {Smith}, {Solano}, {Solitro}, {Sordo}, {Soria Nieto}, {Souchay}, {Spagna}, {Spoto}, {Stampa}, {Steele}, {Steidelm{\"u}ller}, {Stephenson}, {Stoev}, {Suess}, {Surdej}, {Szabados}, {Szegedi-Elek}, {Tapiador}, {Taris}, {Tauran}, {Taylor},
  {Teixeira}, {Terrett}, {Teyssandier}, {Thuillot}, {Titarenko}, {Torra Clotet}, {Turon}, {Ulla}, {Utrilla}, {Uzzi}, {Vaillant}, {Valentini}, {Valette}, {van Elteren}, {Van Hemelryck}, {van Leeuwen}, {Vaschetto}, {Vecchiato}, {Veljanoski}, {Viala}, {Vicente}, {Vogt}, {von Essen}, {Voss}, {Votruba}, {Voutsinas}, {Walmsley}, {Weiler}, {Wertz}, {Wevers}, {Wyrzykowski}, {Yoldas}, {{\v{Z}}erjal}, {Ziaeepour}, {Zorec}, {Zschocke}, {Zucker}, {Zurbach}, \& {Zwitter}}]{Gaia2018}
{Gaia Collaboration}, {Brown}, A.~G.~A., {Vallenari}, A., {et~al.} 2018, \aap, 616, A1

\bibitem[{{G{\'o}mez} {et~al.}(2012){G{\'o}mez}, {Coleman-Smith}, {O'Shea}, {Tumlinson}, \& {Wolpert}}]{Gomez2012}
{G{\'o}mez}, F.~A., {Coleman-Smith}, C.~E., {O'Shea}, B.~W., {Tumlinson}, J., \& {Wolpert}, R.~L. 2012, \apj, 760, 112

\bibitem[{{Gonzalez-Jara} {et~al.}(2025){Gonzalez-Jara}, {Tissera}, {Monachesi}, {Sillero}, {Pallero}, {Pedrosa}, {Tau}, {Tapia-Contreras}, \& {Bignone}}]{Gonzalez-Jara2025}
{Gonzalez-Jara}, J., {Tissera}, P.~B., {Monachesi}, A., {et~al.} 2025, \aap, 693, A282

\bibitem[{{Graus} {et~al.}(2019){Graus}, {Bullock}, {Fitts}, {Cooper}, {Boylan-Kolchin}, {Weisz}, {Wetzel}, {Feldmann}, {Faucher-Gigu{\`e}re}, {Quataert}, {Hopkins}, \& {Kere{\v{s}}}}]{Graus2019}
{Graus}, A.~S., {Bullock}, J.~S., {Fitts}, A., {et~al.} 2019, \mnras, 490, 1186

\bibitem[{{Harmsen} {et~al.}(2017){Harmsen}, {Monachesi}, {Bell}, {de Jong}, {Bailin}, {Radburn-Smith}, \& {Holwerda}}]{Harmsen2017}
{Harmsen}, B., {Monachesi}, A., {Bell}, E.~F., {et~al.} 2017, \mnras, 466, 1491

\bibitem[{{Helmi} {et~al.}(2018){Helmi}, {Babusiaux}, {Koppelman}, {Massari}, {Veljanoski}, \& {Brown}}]{Helmi2018}
{Helmi}, A., {Babusiaux}, C., {Koppelman}, H.~H., {et~al.} 2018, \nat, 563, 85

\bibitem[{{Helmi} \& {White}(1999)}]{Helmi1999}
{Helmi}, A. \& {White}, S. D.~M. 1999, \mnras, 307, 495

\bibitem[{{Helmi} {et~al.}(2003){Helmi}, {White}, \& {Springel}}]{Helmi2003}
{Helmi}, A., {White}, S. D.~M., \& {Springel}, V. 2003, \mnras, 339, 834

\bibitem[{{Ibata} {et~al.}(2014){Ibata}, {Lewis}, {McConnachie}, {Martin}, {Irwin}, {Ferguson}, {Babul}, {Bernard}, {Chapman}, {Collins}, {Fardal}, {Mackey}, {Navarro}, {Pe{\~n}arrubia}, {Rich}, {Tanvir}, \& {Widrow}}]{Ibata2014}
{Ibata}, R.~A., {Lewis}, G.~F., {McConnachie}, A.~W., {et~al.} 2014, \apj, 780, 128

\bibitem[{{Ivezi{\'c}} {et~al.}(2019){Ivezi{\'c}}, {Kahn}, {Tyson}, {Abel}, {Acosta}, {Allsman}, {Alonso}, {AlSayyad}, {Anderson}, {Andrew}, {Angel}, {Angeli}, {Ansari}, {Antilogus}, {Araujo}, {Armstrong}, {Arndt}, {Astier}, {Aubourg}, {Auza}, {Axelrod}, {Bard}, {Barr}, {Barrau}, {Bartlett}, {Bauer}, {Bauman}, {Baumont}, {Bechtol}, {Bechtol}, {Becker}, {Becla}, {Beldica}, {Bellavia}, {Bianco}, {Biswas}, {Blanc}, {Blazek}, {Blandford}, {Bloom}, {Bogart}, {Bond}, {Booth}, {Borgland}, {Borne}, {Bosch}, {Boutigny}, {Brackett}, {Bradshaw}, {Brandt}, {Brown}, {Bullock}, {Burchat}, {Burke}, {Cagnoli}, {Calabrese}, {Callahan}, {Callen}, {Carlin}, {Carlson}, {Chandrasekharan}, {Charles-Emerson}, {Chesley}, {Cheu}, {Chiang}, {Chiang}, {Chirino}, {Chow}, {Ciardi}, {Claver}, {Cohen-Tanugi}, {Cockrum}, {Coles}, {Connolly}, {Cook}, {Cooray}, {Covey}, {Cribbs}, {Cui}, {Cutri}, {Daly}, {Daniel}, {Daruich}, {Daubard}, {Daues}, {Dawson}, {Delgado}, {Dellapenna}, {de Peyster}, {de Val-Borro}, {Digel}, {Doherty}, {Dubois},
  {Dubois-Felsmann}, {Durech}, {Economou}, {Eifler}, {Eracleous}, {Emmons}, {Fausti Neto}, {Ferguson}, {Figueroa}, {Fisher-Levine}, {Focke}, {Foss}, {Frank}, {Freemon}, {Gangler}, {Gawiser}, {Geary}, {Gee}, {Geha}, {Gessner}, {Gibson}, {Gilmore}, {Glanzman}, {Glick}, {Goldina}, {Goldstein}, {Goodenow}, {Graham}, {Gressler}, {Gris}, {Guy}, {Guyonnet}, {Haller}, {Harris}, {Hascall}, {Haupt}, {Hernandez}, {Herrmann}, {Hileman}, {Hoblitt}, {Hodgson}, {Hogan}, {Howard}, {Huang}, {Huffer}, {Ingraham}, {Innes}, {Jacoby}, {Jain}, {Jammes}, {Jee}, {Jenness}, {Jernigan}, {Jevremovi{\'c}}, {Johns}, {Johnson}, {Johnson}, {Jones}, {Juramy-Gilles}, {Juri{\'c}}, {Kalirai}, {Kallivayalil}, {Kalmbach}, {Kantor}, {Karst}, {Kasliwal}, {Kelly}, {Kessler}, {Kinnison}, {Kirkby}, {Knox}, {Kotov}, {Krabbendam}, {Krughoff}, {Kub{\'a}nek}, {Kuczewski}, {Kulkarni}, {Ku}, {Kurita}, {Lage}, {Lambert}, {Lange}, {Langton}, {Le Guillou}, {Levine}, {Liang}, {Lim}, {Lintott}, {Long}, {Lopez}, {Lotz}, {Lupton}, {Lust}, {MacArthur}, {Mahabal},
  {Mandelbaum}, {Markiewicz}, {Marsh}, {Marshall}, {Marshall}, {May}, {McKercher}, {McQueen}, {Meyers}, {Migliore}, {Miller}, {Mills}, {Miraval}, {Moeyens}, {Moolekamp}, {Monet}, {Moniez}, {Monkewitz}, {Montgomery}, {Morrison}, {Mueller}, {Muller}, {Mu{\~n}oz Arancibia}, {Neill}, {Newbry}, {Nief}, {Nomerotski}, {Nordby}, {O'Connor}, {Oliver}, {Olivier}, {Olsen}, {O'Mullane}, {Ortiz}, {Osier}, {Owen}, {Pain}, {Palecek}, {Parejko}, {Parsons}, {Pease}, {Peterson}, {Peterson}, {Petravick}, {Libby Petrick}, {Petry}, {Pierfederici}, {Pietrowicz}, {Pike}, {Pinto}, {Plante}, {Plate}, {Plutchak}, {Price}, {Prouza}, {Radeka}, {Rajagopal}, {Rasmussen}, {Regnault}, {Reil}, {Reiss}, {Reuter}, {Ridgway}, {Riot}, {Ritz}, {Robinson}, {Roby}, {Roodman}, {Rosing}, {Roucelle}, {Rumore}, {Russo}, {Saha}, {Sassolas}, {Schalk}, {Schellart}, {Schindler}, {Schmidt}, {Schneider}, {Schneider}, {Schoening}, {Schumacher}, {Schwamb}, {Sebag}, {Selvy}, {Sembroski}, {Seppala}, {Serio}, {Serrano}, {Shaw}, {Shipsey}, {Sick}, {Silvestri},
  {Slater}, {Smith}, {Smith}, {Sobhani}, {Soldahl}, {Storrie-Lombardi}, {Stover}, {Strauss}, {Street}, {Stubbs}, {Sullivan}, {Sweeney}, {Swinbank}, {Szalay}, {Takacs}, {Tether}, {Thaler}, {Thayer}, {Thomas}, {Thornton}, {Thukral}, {Tice}, {Trilling}, {Turri}, {Van Berg}, {Vanden Berk}, {Vetter}, {Virieux}, {Vucina}, {Wahl}, {Walkowicz}, {Walsh}, {Walter}, {Wang}, {Wang}, {Warner}, {Wiecha}, {Willman}, {Winters}, {Wittman}, {Wolff}, {Wood-Vasey}, {Wu}, {Xin}, {Yoachim}, \& {Zhan}}]{LSST2019}
{Ivezi{\'c}}, {\v{Z}}., {Kahn}, S.~M., {Tyson}, J.~A., {et~al.} 2019, \apj, 873, 111

\bibitem[{{Johnston} {et~al.}(1996){Johnston}, {Hernquist}, \& {Bolte}}]{Johnston1996}
{Johnston}, K.~V., {Hernquist}, L., \& {Bolte}, M. 1996, \apj, 465, 278

\bibitem[{{Kirby} {et~al.}(2013){Kirby}, {Cohen}, {Guhathakurta}, {Cheng}, {Bullock}, \& {Gallazzi}}]{Kirby2013}
{Kirby}, E.~N., {Cohen}, J.~G., {Guhathakurta}, P., {et~al.} 2013, \apj, 779, 102

\bibitem[{{Koppelman} {et~al.}(2019){Koppelman}, {Helmi}, {Massari}, {Roelenga}, \& {Bastian}}]{Koppelman2019}
{Koppelman}, H.~H., {Helmi}, A., {Massari}, D., {Roelenga}, S., \& {Bastian}, U. 2019, \aap, 625, A5

\bibitem[{{Lee} {et~al.}(2006){Lee}, {Skillman}, {Cannon}, {Jackson}, {Gehrz}, {Polomski}, \& {Woodward}}]{Lee2006}
{Lee}, H., {Skillman}, E.~D., {Cannon}, J.~M., {et~al.} 2006, \apj, 647, 970

\bibitem[{{Majewski} {et~al.}(2017){Majewski}, {Schiavon}, {Frinchaboy}, {Allende Prieto}, {Barkhouser}, {Bizyaev}, {Blank}, {Brunner}, {Burton}, {Carrera}, {Chojnowski}, {Cunha}, {Epstein}, {Fitzgerald}, {Garc{\'\i}a P{\'e}rez}, {Hearty}, {Henderson}, {Holtzman}, {Johnson}, {Lam}, {Lawler}, {Maseman}, {M{\'e}sz{\'a}ros}, {Nelson}, {Nguyen}, {Nidever}, {Pinsonneault}, {Shetrone}, {Smee}, {Smith}, {Stolberg}, {Skrutskie}, {Walker}, {Wilson}, {Zasowski}, {Anders}, {Basu}, {Beland}, {Blanton}, {Bovy}, {Brownstein}, {Carlberg}, {Chaplin}, {Chiappini}, {Eisenstein}, {Elsworth}, {Feuillet}, {Fleming}, {Galbraith-Frew}, {Garc{\'\i}a}, {Garc{\'\i}a-Hern{\'a}ndez}, {Gillespie}, {Girardi}, {Gunn}, {Hasselquist}, {Hayden}, {Hekker}, {Ivans}, {Kinemuchi}, {Klaene}, {Mahadevan}, {Mathur}, {Mosser}, {Muna}, {Munn}, {Nichol}, {O'Connell}, {Parejko}, {Robin}, {Rocha-Pinto}, {Schultheis}, {Serenelli}, {Shane}, {Silva Aguirre}, {Sobeck}, {Thompson}, {Troup}, {Weinberg}, \& {Zamora}}]{Majewski2017}
{Majewski}, S.~R., {Schiavon}, R.~P., {Frinchaboy}, P.~M., {et~al.} 2017, \aj, 154, 94

\bibitem[{{Malin} \& {Carter}(1980)}]{MalinCarter81}
{Malin}, D.~F. \& {Carter}, D. 1980, \nat, 285, 643

\bibitem[{{Malin} \& {Carter}(1983)}]{MalinCarter83}
{Malin}, D.~F. \& {Carter}, D. 1983, \apj, 274, 534

\bibitem[{{Marinacci} {et~al.}(2018){Marinacci}, {Vogelsberger}, {Pakmor}, {Torrey}, {Springel}, {Hernquist}, {Nelson}, {Weinberger}, {Pillepich}, {Naiman}, \& {Genel}}]{Miranacci2018}
{Marinacci}, F., {Vogelsberger}, M., {Pakmor}, R., {et~al.} 2018, \mnras, 480, 5113

\bibitem[{{Martin} {et~al.}(2025){Martin}, {Watkins}, {Dubois}, {Devriendt}, {Kaviraj}, {Kim}, {Kraljic}, {Lazar}, {Pearce}, {Peirani}, {Pichon}, {Slyz}, \& {Yi}}]{Martin2025}
{Martin}, G., {Watkins}, A.~E., {Dubois}, Y., {et~al.} 2025, \mnras [\eprint[arXiv]{2505.04509}]

\bibitem[{{Mart{\'\i}nez-Delgado} {et~al.}(2023){Mart{\'\i}nez-Delgado}, {Cooper}, {Rom{\'a}n}, {Pillepich}, {Erkal}, {Pearson}, {Moustakas}, {Laporte}, {Laine}, {Akhlaghi}, {Lang}, {Makarov}, {Borlaff}, {Donatiello}, {Pearson}, {Mir{\'o}-Carretero}, {Cuillandre}, {Dom{\'\i}nguez}, {Roca-F{\`a}brega}, {Frenk}, {Schmidt}, {G{\'o}mez-Flechoso}, {Guzman}, {Libeskind}, {Dey}, {Weaver}, {Schlegel}, {Myers}, \& {Valdes}}]{M-D2023}
{Mart{\'\i}nez-Delgado}, D., {Cooper}, A.~P., {Rom{\'a}n}, J., {et~al.} 2023, \aap, 671, A141

\bibitem[{{Mateu}(2023)}]{Mateu2023}
{Mateu}, C. 2023, \mnras, 520, 5225

\bibitem[{{McConnachie} {et~al.}(2009){McConnachie}, {Irwin}, {Ibata}, {Dubinski}, {Widrow}, {Martin}, {C{\^o}t{\'e}}, {Dotter}, {Navarro}, {Ferguson}, {Puzia}, {Lewis}, {Babul}, {Barmby}, {Bienaym{\'e}}, {Chapman}, {Cockcroft}, {Collins}, {Fardal}, {Harris}, {Huxor}, {Mackey}, {Pe{\~n}arrubia}, {Rich}, {Richer}, {Siebert}, {Tanvir}, {Valls-Gabaud}, \& {Venn}}]{McConnachie2009}
{McConnachie}, A.~W., {Irwin}, M.~J., {Ibata}, R.~A., {et~al.} 2009, \nat, 461, 66

\bibitem[{{Mercado} {et~al.}(2021){Mercado}, {Bullock}, {Boylan-Kolchin}, {Moreno}, {Wetzel}, {El-Badry}, {Graus}, {Fitts}, {Hopkins}, {Faucher-Gigu{\`e}re}, \& {Gurvich}}]{Mercado2021}
{Mercado}, F.~J., {Bullock}, J.~S., {Boylan-Kolchin}, M., {et~al.} 2021, \mnras, 501, 5121

\bibitem[{{Merritt} {et~al.}(2020){Merritt}, {Pillepich}, {van Dokkum}, {Nelson}, {Hernquist}, {Marinacci}, \& {Vogelsberger}}]{Merritt2020}
{Merritt}, A., {Pillepich}, A., {van Dokkum}, P., {et~al.} 2020, \mnras, 495, 4570

\bibitem[{{Merritt} {et~al.}(2016){Merritt}, {van Dokkum}, {Abraham}, \& {Zhang}}]{Merritt2016}
{Merritt}, A., {van Dokkum}, P., {Abraham}, R., \& {Zhang}, J. 2016, \apj, 830, 62

\bibitem[{{Monachesi} {et~al.}(2016){Monachesi}, {Bell}, {Radburn-Smith}, {Bailin}, {de Jong}, {Holwerda}, {Streich}, \& {Silverstein}}]{Monachesi2016}
{Monachesi}, A., {Bell}, E.~F., {Radburn-Smith}, D.~J., {et~al.} 2016, \mnras, 457, 1419

\bibitem[{{Monachesi} {et~al.}(2019){Monachesi}, {G{\'o}mez}, {Grand}, {Simpson}, {Kauffmann}, {Bustamante}, {Marinacci}, {Pakmor}, {Springel}, {Frenk}, {White}, \& {Tissera}}]{Monachesi2019}
{Monachesi}, A., {G{\'o}mez}, F.~A., {Grand}, R. J.~J., {et~al.} 2019, \mnras, 485, 2589

\bibitem[{{Moster} {et~al.}(2018){Moster}, {Naab}, \& {White}}]{Moster2018}
{Moster}, B.~P., {Naab}, T., \& {White}, S. D.~M. 2018, \mnras, 477, 1822

\bibitem[{{Naab} \& {Burkert}(2003)}]{Naab2003}
{Naab}, T. \& {Burkert}, A. 2003, \apj, 597, 893

\bibitem[{{Naiman} {et~al.}(2018){Naiman}, {Pillepich}, {Springel}, {Ramirez-Ruiz}, {Torrey}, {Vogelsberger}, {Pakmor}, {Nelson}, {Marinacci}, {Hernquist}, {Weinberger}, \& {Genel}}]{Naiman2018}
{Naiman}, J.~P., {Pillepich}, A., {Springel}, V., {et~al.} 2018, \mnras, 477, 1206

\bibitem[{{Navarro} {et~al.}(1997){Navarro}, {Frenk}, \& {White}}]{Navarro1997}
{Navarro}, J.~F., {Frenk}, C.~S., \& {White}, S. D.~M. 1997, \apj, 490, 493

\bibitem[{{Navarro} {et~al.}(2018){Navarro}, {Yozin}, {Loewen}, {Ben{\'\i}tez-Llambay}, {Fattahi}, {Frenk}, {Oman}, {Schaye}, \& {Theuns}}]{Navarro2018}
{Navarro}, J.~F., {Yozin}, C., {Loewen}, N., {et~al.} 2018, \mnras, 476, 3648

\bibitem[{{Nelson} {et~al.}(2019{\natexlab{a}}){Nelson}, {Pillepich}, {Springel}, {Pakmor}, {Weinberger}, {Genel}, {Torrey}, {Vogelsberger}, {Marinacci}, \& {Hernquist}}]{Nelson2019TNG50}
{Nelson}, D., {Pillepich}, A., {Springel}, V., {et~al.} 2019{\natexlab{a}}, \mnras, 490, 3234

\bibitem[{{Nelson} {et~al.}(2019{\natexlab{b}}){Nelson}, {Springel}, {Pillepich}, {Rodriguez-Gomez}, {Torrey}, {Genel}, {Vogelsberger}, {Pakmor}, {Marinacci}, {Weinberger}, {Kelley}, {Lovell}, {Diemer}, \& {Hernquist}}]{Nelson2019}
{Nelson}, D., {Springel}, V., {Pillepich}, A., {et~al.} 2019{\natexlab{b}}, Computational Astrophysics and Cosmology, 6, 2

\bibitem[{{Pakmor} {et~al.}(2016){Pakmor}, {Springel}, {Bauer}, {Mocz}, {Munoz}, {Ohlmann}, {Schaal}, \& {Zhu}}]{Pakmor2016}
{Pakmor}, R., {Springel}, V., {Bauer}, A., {et~al.} 2016, \mnras, 455, 1134

\bibitem[{{Pillepich} {et~al.}(2018{\natexlab{a}}){Pillepich}, {Nelson}, {Hernquist}, {Springel}, {Pakmor}, {Torrey}, {Weinberger}, {Genel}, {Naiman}, {Marinacci}, \& {Vogelsberger}}]{Pillepich2018}
{Pillepich}, A., {Nelson}, D., {Hernquist}, L., {et~al.} 2018{\natexlab{a}}, \mnras, 475, 648

\bibitem[{{Pillepich} {et~al.}(2019){Pillepich}, {Nelson}, {Springel}, {Pakmor}, {Torrey}, {Weinberger}, {Vogelsberger}, {Marinacci}, {Genel}, {van der Wel}, \& {Hernquist}}]{Pillepich2019}
{Pillepich}, A., {Nelson}, D., {Springel}, V., {et~al.} 2019, \mnras, 490, 3196

\bibitem[{{Pillepich} {et~al.}(2018{\natexlab{b}}){Pillepich}, {Springel}, {Nelson}, {Genel}, {Naiman}, {Pakmor}, {Hernquist}, {Torrey}, {Vogelsberger}, {Weinberger}, \& {Marinacci}}]{Pillepich2018modelTNG}
{Pillepich}, A., {Springel}, V., {Nelson}, D., {et~al.} 2018{\natexlab{b}}, \mnras, 473, 4077

\bibitem[{{Pillepich} {et~al.}(2014){Pillepich}, {Vogelsberger}, {Deason}, {Rodriguez-Gomez}, {Genel}, {Nelson}, {Torrey}, {Sales}, {Marinacci}, {Springel}, {Sijacki}, \& {Hernquist}}]{Pill2014}
{Pillepich}, A., {Vogelsberger}, M., {Deason}, A., {et~al.} 2014, \mnras, 444, 237

\bibitem[{{Planck Collaboration} {et~al.}(2016){Planck Collaboration}, {Ade}, {Aghanim}, {Arnaud}, {Ashdown}, {Aumont}, {Baccigalupi}, {Banday}, {Barreiro}, {Bartlett}, {Bartolo}, {Battaner}, {Battye}, {Benabed}, {Beno{\^\i}t}, {Benoit-L{\'e}vy}, {Bernard}, {Bersanelli}, {Bielewicz}, {Bock}, {Bonaldi}, {Bonavera}, {Bond}, {Borrill}, {Bouchet}, {Boulanger}, {Bucher}, {Burigana}, {Butler}, {Calabrese}, {Cardoso}, {Catalano}, {Challinor}, {Chamballu}, {Chary}, {Chiang}, {Chluba}, {Christensen}, {Church}, {Clements}, {Colombi}, {Colombo}, {Combet}, {Coulais}, {Crill}, {Curto}, {Cuttaia}, {Danese}, {Davies}, {Davis}, {de Bernardis}, {de Rosa}, {de Zotti}, {Delabrouille}, {D{\'e}sert}, {Di Valentino}, {Dickinson}, {Diego}, {Dolag}, {Dole}, {Donzelli}, {Dor{\'e}}, {Douspis}, {Ducout}, {Dunkley}, {Dupac}, {Efstathiou}, {Elsner}, {En{\ss}lin}, {Eriksen}, {Farhang}, {Fergusson}, {Finelli}, {Forni}, {Frailis}, {Fraisse}, {Franceschi}, {Frejsel}, {Galeotta}, {Galli}, {Ganga}, {Gauthier}, {Gerbino}, {Ghosh}, {Giard},
  {Giraud-H{\'e}raud}, {Giusarma}, {Gjerl{\o}w}, {Gonz{\'a}lez-Nuevo}, {G{\'o}rski}, {Gratton}, {Gregorio}, {Gruppuso}, {Gudmundsson}, {Hamann}, {Hansen}, {Hanson}, {Harrison}, {Helou}, {Henrot-Versill{\'e}}, {Hern{\'a}ndez-Monteagudo}, {Herranz}, {Hildebrandt}, {Hivon}, {Hobson}, {Holmes}, {Hornstrup}, {Hovest}, {Huang}, {Huffenberger}, {Hurier}, {Jaffe}, {Jaffe}, {Jones}, {Juvela}, {Keih{\"a}nen}, {Keskitalo}, {Kisner}, {Kneissl}, {Knoche}, {Knox}, {Kunz}, {Kurki-Suonio}, {Lagache}, {L{\"a}hteenm{\"a}ki}, {Lamarre}, {Lasenby}, {Lattanzi}, {Lawrence}, {Leahy}, {Leonardi}, {Lesgourgues}, {Levrier}, {Lewis}, {Liguori}, {Lilje}, {Linden-V{\o}rnle}, {L{\'o}pez-Caniego}, {Lubin}, {Mac{\'\i}as-P{\'e}rez}, {Maggio}, {Maino}, {Mandolesi}, {Mangilli}, {Marchini}, {Maris}, {Martin}, {Martinelli}, {Mart{\'\i}nez-Gonz{\'a}lez}, {Masi}, {Matarrese}, {McGehee}, {Meinhold}, {Melchiorri}, {Melin}, {Mendes}, {Mennella}, {Migliaccio}, {Millea}, {Mitra}, {Miville-Desch{\^e}nes}, {Moneti}, {Montier}, {Morgante}, {Mortlock},
  {Moss}, {Munshi}, {Murphy}, {Naselsky}, {Nati}, {Natoli}, {Netterfield}, {N{\o}rgaard-Nielsen}, {Noviello}, {Novikov}, {Novikov}, {Oxborrow}, {Paci}, {Pagano}, {Pajot}, {Paladini}, {Paoletti}, {Partridge}, {Pasian}, {Patanchon}, {Pearson}, {Perdereau}, {Perotto}, {Perrotta}, {Pettorino}, {Piacentini}, {Piat}, {Pierpaoli}, {Pietrobon}, {Plaszczynski}, {Pointecouteau}, {Polenta}, {Popa}, {Pratt}, {Pr{\'e}zeau}, {Prunet}, {Puget}, {Rachen}, {Reach}, {Rebolo}, {Reinecke}, {Remazeilles}, {Renault}, {Renzi}, {Ristorcelli}, {Rocha}, {Rosset}, {Rossetti}, {Roudier}, {Rouill{\'e} d'Orfeuil}, {Rowan-Robinson}, {Rubi{\~n}o-Mart{\'\i}n}, {Rusholme}, {Said}, {Salvatelli}, {Salvati}, {Sandri}, {Santos}, {Savelainen}, {Savini}, {Scott}, {Seiffert}, {Serra}, {Shellard}, {Spencer}, {Spinelli}, {Stolyarov}, {Stompor}, {Sudiwala}, {Sunyaev}, {Sutton}, {Suur-Uski}, {Sygnet}, {Tauber}, {Terenzi}, {Toffolatti}, {Tomasi}, {Tristram}, {Trombetti}, {Tucci}, {Tuovinen}, {T{\"u}rler}, {Umana}, {Valenziano}, {Valiviita}, {Van Tent},
  {Vielva}, {Villa}, {Wade}, {Wandelt}, {Wehus}, {White}, {White}, {Wilkinson}, {Yvon}, {Zacchei}, \& {Zonca}}]{Planck2016}
{Planck Collaboration}, {Ade}, P.~A.~R., {Aghanim}, N., {et~al.} 2016, \aap, 594, A13

\bibitem[{{Proctor} {et~al.}(2024){Proctor}, {Ludlow}, {Lagos}, \& {Robotham}}]{Proctor2024}
{Proctor}, K.~L., {Ludlow}, A.~D., {Lagos}, C. d.~P., \& {Robotham}, A. S.~G. 2024, arXiv e-prints, arXiv:2407.11444

\bibitem[{{Purcell} {et~al.}(2007){Purcell}, {Bullock}, \& {Zentner}}]{Purcell2007}
{Purcell}, C.~W., {Bullock}, J.~S., \& {Zentner}, A.~R. 2007, \apj, 666, 20

\bibitem[{{Rich} {et~al.}(2019){Rich}, {Mosenkov}, {Lee-Saunders}, {Koch}, {Kormendy}, {Kennefick}, {Brosch}, {Sales}, {Bullock}, {Burkert}, {Collins}, {Cooper}, {Fusco}, {Reitzel}, {Thilker}, {Milewski}, {Elias}, {Saade}, \& {De Groot}}]{Rich2019}
{Rich}, R.~M., {Mosenkov}, A., {Lee-Saunders}, H., {et~al.} 2019, \mnras, 490, 1539

\bibitem[{{Rodriguez-Gomez} {et~al.}(2015){Rodriguez-Gomez}, {Genel}, {Vogelsberger}, {Sijacki}, {Pillepich}, {Sales}, {Torrey}, {Snyder}, {Nelson}, {Springel}, {Ma}, \& {Hernquist}}]{R-G2015}
{Rodriguez-Gomez}, V., {Genel}, S., {Vogelsberger}, M., {et~al.} 2015, \mnras, 449, 49

\bibitem[{{Rodriguez-Gomez} {et~al.}(2016){Rodriguez-Gomez}, {Pillepich}, {Sales}, {Genel}, {Vogelsberger}, {Zhu}, {Wellons}, {Nelson}, {Torrey}, {Springel}, {Ma}, \& {Hernquist}}]{R-G2016}
{Rodriguez-Gomez}, V., {Pillepich}, A., {Sales}, L.~V., {et~al.} 2016, \mnras, 458, 2371

\bibitem[{{Searle} \& {Zinn}(1978)}]{SearleZinn1978}
{Searle}, L. \& {Zinn}, R. 1978, \apj, 225, 357

\bibitem[{{S{\'e}rsic}(1963)}]{Sersic1963}
{S{\'e}rsic}, J.~L. 1963, Boletin de la Asociacion Argentina de Astronomia La Plata Argentina, 6, 41

\bibitem[{{Springel}(2010)}]{Springel2010}
{Springel}, V. 2010, \mnras, 401, 791

\bibitem[{{Springel} \& {Hernquist}(2003)}]{SpringelHernquist2003}
{Springel}, V. \& {Hernquist}, L. 2003, \mnras, 339, 289

\bibitem[{{Springel} {et~al.}(2018){Springel}, {Pakmor}, {Pillepich}, {Weinberger}, {Nelson}, {Hernquist}, {Vogelsberger}, {Genel}, {Torrey}, {Marinacci}, \& {Naiman}}]{Springel2018}
{Springel}, V., {Pakmor}, R., {Pillepich}, A., {et~al.} 2018, \mnras, 475, 676

\bibitem[{{Springel} {et~al.}(2005){Springel}, {White}, {Jenkins}, {Frenk}, {Yoshida}, {Gao}, {Navarro}, {Thacker}, {Croton}, {Helly}, {Peacock}, {Cole}, {Thomas}, {Couchman}, {Evrard}, {Colberg}, \& {Pearce}}]{Springel2005}
{Springel}, V., {White}, S. D.~M., {Jenkins}, A., {et~al.} 2005, \nat, 435, 629

\bibitem[{{Springel} {et~al.}(2001){Springel}, {White}, {Tormen}, \& {Kauffmann}}]{Springel2001}
{Springel}, V., {White}, S. D.~M., {Tormen}, G., \& {Kauffmann}, G. 2001, \mnras, 328, 726

\bibitem[{{Starkenburg} {et~al.}(2017){Starkenburg}, {Oman}, {Navarro}, {Crain}, {Fattahi}, {Frenk}, {Sawala}, \& {Schaye}}]{Starkenburg2017}
{Starkenburg}, E., {Oman}, K.~A., {Navarro}, J.~F., {et~al.} 2017, \mnras, 465, 2212

\bibitem[{{Tau} {et~al.}(2024){Tau}, {Monachesi}, {Gomez}, {Grand}, {Pakmor}, {van de Voort}, {Gonzalez-Jara}, {Tissera}, {Marinacci}, \& {Bieri}}]{Tau2024}
{Tau}, E.~A., {Monachesi}, A., {Gomez}, F.~A., {et~al.} 2024, arXiv e-prints, arXiv:2412.13807

\bibitem[{{Tissera} {et~al.}(2014){Tissera}, {Beers}, {Carollo}, \& {Scannapieco}}]{Tissera2014}
{Tissera}, P.~B., {Beers}, T.~C., {Carollo}, D., \& {Scannapieco}, C. 2014, \mnras, 439, 3128

\bibitem[{{Tissera} {et~al.}(2013){Tissera}, {Scannapieco}, {Beers}, \& {Carollo}}]{Tissera2013}
{Tissera}, P.~B., {Scannapieco}, C., {Beers}, T.~C., \& {Carollo}, D. 2013, \mnras, 432, 3391

\bibitem[{{Tissera} {et~al.}(2012){Tissera}, {White}, \& {Scannapieco}}]{Tissera2012}
{Tissera}, P.~B., {White}, S. D.~M., \& {Scannapieco}, C. 2012, \mnras, 420, 255

\bibitem[{{Vogelsberger} {et~al.}(2013){Vogelsberger}, {Genel}, {Sijacki}, {Torrey}, {Springel}, \& {Hernquist}}]{Vogelsberger2013}
{Vogelsberger}, M., {Genel}, S., {Sijacki}, D., {et~al.} 2013, \mnras, 436, 3031

\bibitem[{{Vogelsberger} {et~al.}(2014){Vogelsberger}, {Genel}, {Springel}, {Torrey}, {Sijacki}, {Xu}, {Snyder}, {Bird}, {Nelson}, \& {Hernquist}}]{Vogelsberger2014}
{Vogelsberger}, M., {Genel}, S., {Springel}, V., {et~al.} 2014, \nat, 509, 177

\bibitem[{{Weinberger} {et~al.}(2017){Weinberger}, {Springel}, {Hernquist}, {Pillepich}, {Marinacci}, {Pakmor}, {Nelson}, {Genel}, {Vogelsberger}, {Naiman}, \& {Torrey}}]{Weinberger2017}
{Weinberger}, R., {Springel}, V., {Hernquist}, L., {et~al.} 2017, \mnras, 465, 3291

\bibitem[{{White}(1982)}]{White1982}
{White}, S.~D.~M. 1982, Saas-Fee Advanced Course, 12, 291

\bibitem[{{White} \& {Rees}(1978)}]{WhiteRees1978}
{White}, S.~D.~M. \& {Rees}, M.~J. 1978, \mnras, 183, 341

\end{thebibliography}

\end{document}